\documentclass[%
 reprint,
 superscriptaddress,
 amsmath,amssymb,
 aps,
prb,
]{revtex4-2}

\usepackage{xfrac}
\usepackage{graphicx}
\usepackage{dcolumn}
\usepackage{bm}
\usepackage{amssymb}
\usepackage{amsthm}
\usepackage[version=4]{mhchem}
\usepackage{subcaption} 
\usepackage{float} 
\usepackage{xcolor}
\usepackage{comment}
\usepackage{newunicodechar}
\newunicodechar{−}{\ensuremath{-}}
\usepackage{soulutf8}
\setstcolor{red}   

\begin{document}
\preprint{APS/123-QED}

\title{Influence of Ni substitution on the phase transitions and magnetocaloric effect of \ce{NdCo2} at cryogenic temperatures}

\author{Vilde G. S. Lunde}\thanks{Contact author: vilde.lunde@ife.no}
\affiliation{Department for Hydrogen Technology, Institute for Energy Technology (IFE), PO Box 40, NO-2027, Kjeller, Norway}
\author{{\O}ystein S. Fjellv{\aa}g}
\affiliation{Department for Hydrogen Technology, Institute for Energy Technology (IFE), PO Box 40, NO-2027, Kjeller, Norway}
\author{Allan M. Döring}
\affiliation{Functional Materials, Institute of Materials Science, Technical University of Darmstadt, Darmstadt, 64287, Germany}
\author{Marc Stra{\ss}heim}
\affiliation{Dresden High Magnetic Field Laboratory (HLD-EMFL) and W\"urzburg-Dresden Cluster of Excellence ct.qmat, Helmholtz-Zentrum Dresden-Rossendorf, 01328 Dresden, Germany}
\affiliation{Institut für Festkörper- und Materialphysik, Technische Universität Dresden, 01069 Dresden, Germany}
\author{Vladimir Pomjakushin}
\affiliation{Paul Scherrer Institute (PSI), 5232, Villigen, Switzerland}
\author{Konstantin P. Skokov}
\affiliation{Functional Materials, Institute of Materials Science, Technical University of Darmstadt, Darmstadt, 64287, Germany}
\author{Oliver Gutfleisch}
\affiliation{Functional Materials, Institute of Materials Science, Technical University of Darmstadt, Darmstadt, 64287, Germany}
\author{Tino Gottschall}
\affiliation{Dresden High Magnetic Field Laboratory (HLD-EMFL) and W\"urzburg-Dresden Cluster of Excellence ct.qmat, Helmholtz-Zentrum Dresden-Rossendorf, 01328 Dresden, Germany}
\author{Joachim Wosnitza}
\affiliation{Dresden High Magnetic Field Laboratory (HLD-EMFL) and W\"urzburg-Dresden Cluster of Excellence ct.qmat, Helmholtz-Zentrum Dresden-Rossendorf, 01328 Dresden, Germany}
\affiliation{Institut für Festkörper- und Materialphysik, Technische Universität Dresden, 01069 Dresden, Germany}
\author{Anja O. Sj{\aa}stad}
\affiliation{Department of Chemistry and Centre for Material Science and Nanotechnology (SMN), University of Oslo, NO-0315, Norway}
\author{Bj{\o}rn C. Hauback}
\affiliation{Department for Hydrogen Technology, Institute for Energy Technology (IFE), PO Box 40, NO-2027, Kjeller, Norway}
\author{Christoph Frommen} \thanks{Contact author: christoph.frommen@ife.no}
\affiliation{Department for Hydrogen Technology, Institute for Energy Technology (IFE), PO Box 40, NO-2027, Kjeller, Norway}

\date{\today}

\begin{abstract} 
We have investigated \ce{NdCo_{2-x}Ni_x} cubic Laves compounds with 0 $\leq$ x $\leq$ 1 using neutron diffraction and bulk magnetization measurements to study the influence of partial Ni substitutions of Co on the phase transitions and the magnetocaloric effect. Upon cooling, \ce{NdCo2} undergoes a cubic to tetragonal transition at 100~K, and a tetragonal to orthorhombic transition at 42~K. The transitions are associated with long-range ferromagnetic ordering of the magnetic moments along the $c$ axis and spin reorientation into the $ab$ plane, respectively. Both transitions shift to lower temperatures as the Ni content $x$ increases. For $x \geq 0.5$, the orthorhombic phase is suppressed. Additionally, there was a reduction in the magnetic moment upon increasing the Ni substitution of Co.
The magnetocaloric effect was determined both indirectly and directly, with good agreement between the methods. \ce{NdCo2} exhibits an adiabatic temperature change of 6.3~K for a field of 20~T, which is decreased to 4.9~K for NdCoNi for the same field strength due to the reduced magnetic moment upon Ni substitution.

\end{abstract} 

\maketitle

\section{Introduction}
\label{sec:Introduction} 

Magnetic refrigeration utilizing the magnetocaloric effect (MCE) offers superior cooling efficiency at cryogenic temperatures (20-77~K) compared to conventional gas compression-expansion methods~\cite{franco_magnetocaloric_2018}. Consequently, significant efforts have been made to discover materials with a large MCE in this temperature region. Laves compounds are a class of materials that have shown significant MCEs at low temperatures, but most high-performing candidates contain expensive and scarce heavy rare-earth elements with high magnetic moments, such as Ho, Gd, and Dy~\cite{castro_machine-learning-guided_2020,tereshina_effects_2023, franco_magnetocaloric_2018}. Therefore, there is a need to identify more cost-effective and abundant materials to scale up the technology. Replacing the heavy rare-earth elements in Laves compounds with more abundant light rare-earth elements at the expense of a lower MCE is one promising strategy to reduce cost~\cite{liu_matter_2024}.

\ce{NdCo2} is a well-studied light rare-earth cubic \ce{AB2} Laves phase (space group $Fd\bar{3}m$) which is paramagnetic at room temperature. The compound exhibits a magnetostructural phase transition at 100~K associated with a tetragonal distortion (space group $I4_1/amd$) and ferromagnetic ordering ($T_\text{C}$). The nature of this phase transition has been a subject of some debate in the literature. Most studies agree that the phase transition is of a second-order nature~\cite{xiao_crystal_2006, murtaza_magnetocaloric_2020, dublon_crystal_1976, atzmony_crystal-field-induced_1976, ouyang_magnetic_2005, herrero2006nature}. However, studies based on perturbed angular correlation~\cite{PhysRevB.68.014409, forker2007comment} and resonant ultrasound spectroscopy~\cite{driver2014multiferroic} argue that the transition is first-order or close to tricritical in character. This discrepancy highlights the challenge of definitively characterizing magnetostructural transitions involving coupled magnetic and structural order parameters. Additionally, \ce{NdCo2} undergoes a first-order phase transition at 42~K associated with an orthorhombic distortion (space group $Imma$) and a spin reorientation ($T_\text{SR}$)~\cite{xiao_crystal_2006, murtaza_magnetocaloric_2020, dublon_crystal_1976, atzmony_crystal-field-induced_1976, ouyang_magnetic_2005}. 
Since Co is a resource-critical metal, raising ethical and environmental concerns~\cite{european_commision}, partial substitution with other elements, such as Ni, can reduce the criticality of the compound. Our previous investigations of NdCoNi have revealed that the compound exhibits a $T_\text{C}$ at 36~K, but no $T_\text{SR}$~\cite{lunde_machine_2025, lunde_electronic_2025}. 

First-order magnetic transitions can exhibit high MCEs but are often accompanied by thermal hysteresis~\cite{law_quantitative_2018}. In contrast, second-order magnetic transitions typically result in both lower thermal hysteresis and MCE. The MCE is quantified by the magnetic entropy change $\Delta S_\text{m}$ and the adiabatic temperature change $\Delta T_\text{ad}$. To distinguish between the indirectly and directly determined $\Delta T_\text{ad}$ in this work, we will denote these $\Delta T_\text{ad,ind}$ and $\Delta T_\text{ad,dir}$, respectively. $\Delta T_\text{ad,ind}$ can either be determined from $S$-$T$-diagrams constructed from heat capacity $C_\text{P}$ data, or as $\Delta T_\text{ad,ind} = -\Delta S_\text{m}(T/C_\text{P})$ where $\Delta S_\text{m}$ is determined from field-dependent magnetization isotherms $M(H)$. The latter method can be unreliable for first-order phase transitions and at low temperatures, where $C_\text{P}$ approaches zero, and should be applied with caution to avoid an area of phase coexistence. $\Delta T_\text{ad,dir}$ can be determined by measuring the temperature change under application of a magnetic field. This direct $\Delta T_\text{ad,dir}$ is generally more reliable than the indirect $\Delta T_\text{ad,ind}$, in particular for first-order magnetic transitions~\cite{politova_magnetism_2024}. Pulsed-field measurements can be used to accurately determine $\Delta T_\text{ad,dir}$ by ensuring adiabatic conditions in the experiment~\cite{salazar_mejia_high-field_2023}. 

In this work, \ce{NdCo_{2-x}Ni_x} with (0 $\leq$ x $\leq$ 1) were studied to investigate the influence of Ni substitution on the MCE. The crystal structures, as well as local and bulk magnetic moments, were investigated in detail to understand how these properties are influenced by the partial substitution of Co with Ni. Studying the magnetic properties of a material is essential to understand the origin of its MCE. Therefore, neutron scattering was used to probe both magnetic and crystal structures of the compounds. These experiments are important for understanding the relationship between crystal structure, magnetic ordering, and MCE, to optimize the material performance.

\section{Methods}
\label{sec:Methods} 

Five \ce{NdCo_{2-x}Ni_x} ($x=$ 0, 0.25, 0.50, 0.75, 1) compositions ($\sim$1~g batches) were synthesized by arc melting (Edmund Bühler, Compact Arc Melter MAM-1) under argon atmosphere ($\geq$ 99.99~\%, Nippon Gases) from high-purity elements ($\geq$ 99~\%, Goodfellow), with 3~wt.\% excess of Nd to compensate for evaporation losses~\cite{macaluso_challenges_2012}. The chamber was evacuated and purged with argon gas three times, and left under a final pressure of 0.7~bar. The ingots were turned and remelted four times, followed by 72 hours of annealing at 1073~K and 72 hours at 833~K in evacuated and sealed quartz tubes to improve homogeneity. The final products were cooled to room temperature at a rate of 2~K/min and stored under argon atmosphere to prevent oxidation.

Microstructure and chemical compositions were characterized using a JEOL JSM-7900F scanning electron microscope (SEM) equipped with an Oxford Instruments AZtec Ultim Max 65 energy-dispersive x-ray (EDS) analyzer. Bulk pieces of the samples were embedded in "polyfast" resin, ground, and polished. Back-scattered electron (BSE) mode was utilized to obtain elemental contrast, and EDS mode was used for elemental analysis, both using an acceleration voltage of 15~kV.

Synchrotron radiation powder x-ray diffraction (SR-PXD) data were obtained at BM01 at the Swiss-Norwegian Beamlines (SNBL) of the European Synchrotron Radiation Facility (ESRF) in Grenoble, France. Powder samples were prepared in 0.3~mm quartz capillaries and measured using the PILATUS2M detector~\cite{SNBL} while rotating the capillary 360 degrees. All samples were measured at room temperature with a wavelength of 0.49537~{\AA}. Additionally, \ce{NdCo2} was cooled to 80~K using a nitrogen blower, and measured with a wavelength of 0.71859~{\AA} while heating with a rate of 1~K/min to 120~K, followed by 5~K/min to 300~K. The data were integrated with the SNBL Bubble software~\cite{SNBL}, and Rietveld refinements were performed using Topas V5~\cite{topas}. 

Powder neutron diffraction (PND) measurements were performed using the high-resolution powder diffractometer for thermal neutrons (HRPT) at the Swiss Spallation Neutron Source (SINQ) at the Paul Scherrer Institute (PSI), Villigen, Switzerland. Unless otherwise stated, measurements were performed using a wavelength of 1.886~{\AA} and the high-intensity mode (40' primary beam collimator). Measurements were performed between 1.5 and 120~K. At 1.5~K, additional measurements were performed using the medium-resolution mode (12' primary beam collimator), and at a wavelength of 1.154~{\AA} to determine the thermal displacement parameter. The medium-resolution measurements were also performed at 45~K for \ce{NdCo2} and \ce{NdCo_{1.75}Ni_{0.25}}.
Refinements were performed using Mag2Pol~\cite{mag2pol}, and the magnetic symmetry analysis was done using ISODISTORT~\cite{stokes_isodistort, campbell_isodisplace}.

Magnetic and magnetocaloric properties were characterized using a Physical Property Measurement System (PPMS). 
Temperature-dependent magnetization was measured on zero-field-cooled samples during heating and cooling at a rate of 2~K/min, with applied fields of 0.01, 1, and 5~T, respectively. 
Hysteresis loops were obtained at 5~K between $-$5 and +5~T with a magnetic field-sweep rate of 0.3~T/min. 
$C_\text{P}$ measurements were performed during heating at 0 and 5~T, respectively, using the $2\tau$ approach. $\Delta S_\text{m}$ and $\Delta T_\text{ad,ind}$ were determined from the vertical and horizontal difference between the 0 and 5~T curves in an $S$-$T$-diagram, respectively.
Additionally, $M(H)$ isotherms were measured from 0 to 5~T with 4~K intervals in the temperature region of interest to calculate $\Delta S_\text{m}$ for comparison with the $C_\text{P}$-based results. 
A comprehensive description of the $C_\text{P}$ and $M(H)$ methodologies is provided in the Supplementary Material~\cite{supplemental}.

$\Delta T_\text{ad,dir}$ was measured at the Dresden High Magnetic Field Laboratory (HLD) at Helmholtz-Zentrum Dresden-Rossendorf (HZDR), Germany. Measurements were performed between 10 and 120~K for pulsed magnetic fields of 5, 10, and 20~T. For a comprehensive description of the experimental setup, see Ref.~\cite{salazar_mejia_high-field_2023}.

\section{Results}
\label{sec:Results} 

\subsection{Structural properties and phase transitions}

Initially, the samples were characterized with SR-PXD and SEM-EDS at room temperature. All five compositions of the \ce{NdCo_{2-x}Ni_x} ($x=$ 0, 0.25, 0.50, 0.75, 1) series were found to be of high purity, with minor secondary phases. This was quantified by Rietveld refinements and SEM-EDS analysis, see the Supplemental Material for details~\cite{supplemental}. Both methods primarily show the C15 cubic \ce{AB2} Laves phase, but identify small amounts of \ce{AB}, \ce{AB3}, or oxide impurities (1.0-3.3~wt.\%). Note that none of these phases were identified by PND (see below).

\begin{figure*}[htbp] 
    \centering
    \includegraphics[width=0.49\textwidth]{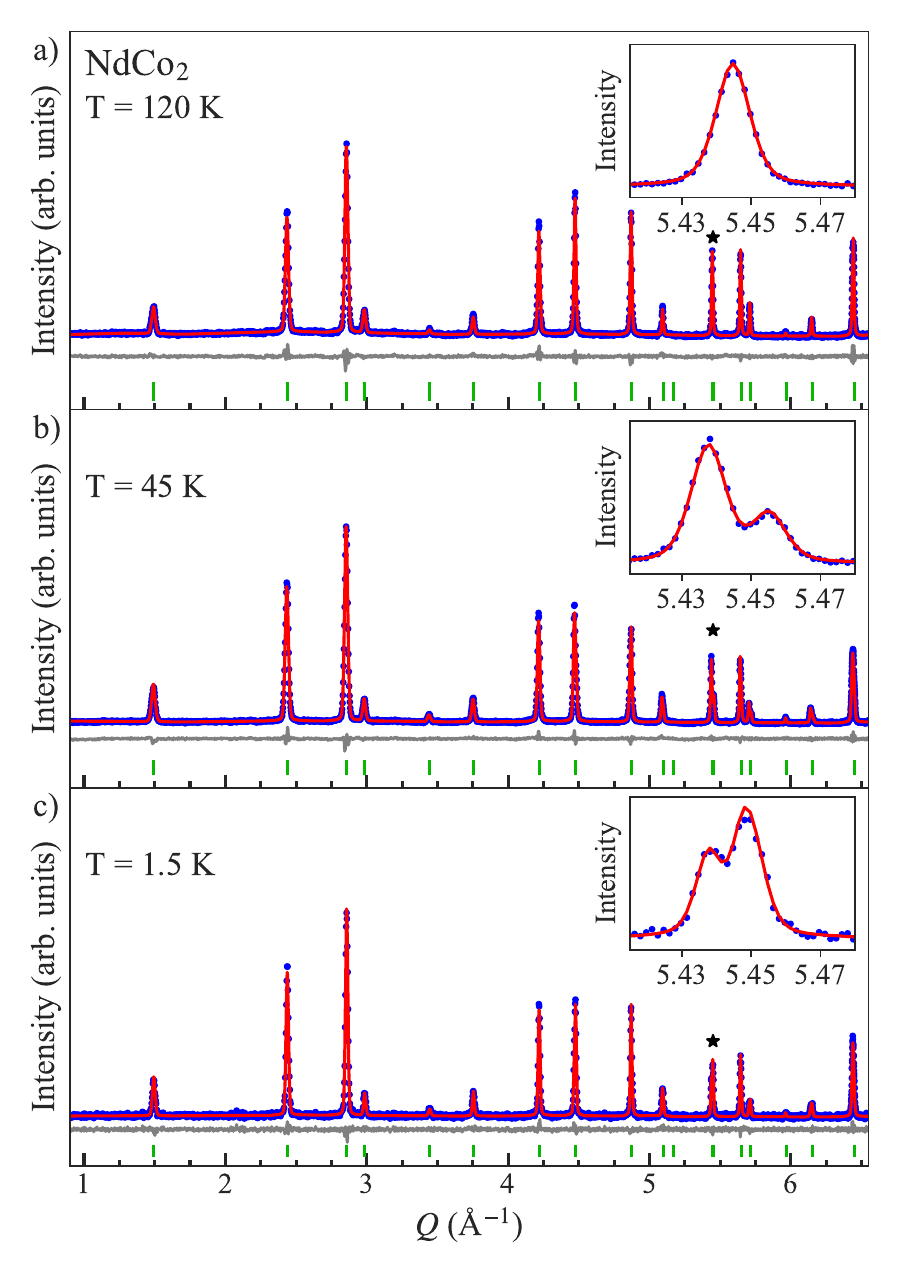}
    \includegraphics[width=0.49\textwidth]{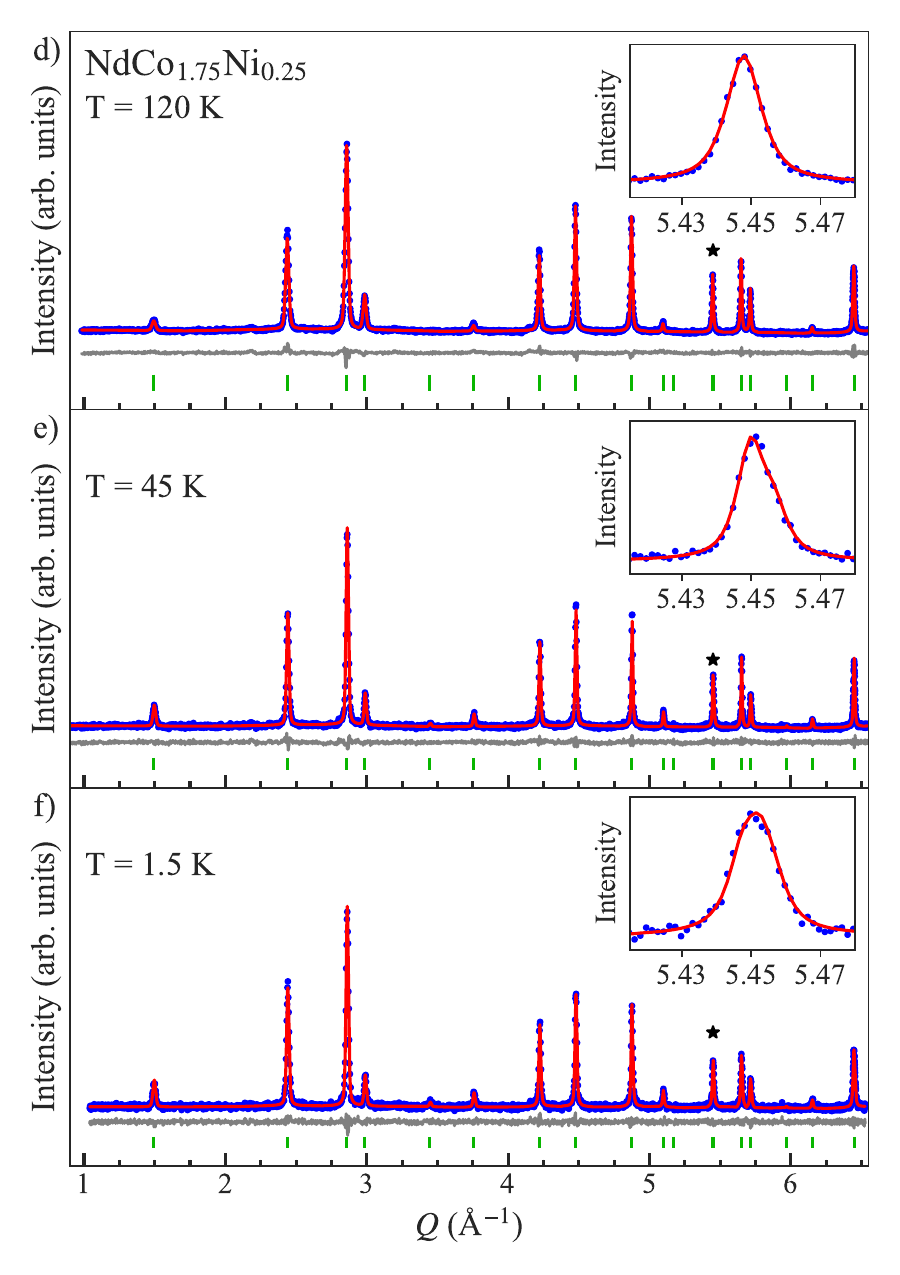}
    \caption{Experimental data (blue circles), calculated diffraction pattern (red lines), and difference (grey lines) for the Rietveld refinements of the PND patterns collected with $\lambda = 1.886$~{\AA} of \ce{NdCo2} (a)-(c) and \ce{NdCo_{1.75}Ni_{0.25}} (d)-(f) of the cubic (120~K, space group $Fd\bar{3}m$), tetragonal (45~K, space group $I4_1/amd$), and orthorhombic (1.5~K, space group $Imma$) phases, respectively. Error bars are within the size of the markers. Vertical bars indicate Bragg reflections for the refined \ce{AB2} phase. The inset shows the (620) Bragg peak, marked with a star in the plot. Rietveld refinements for \ce{NdCo_{1.50}Ni_{0.50}}, \ce{NdCo_{1.25}Ni_{0.75}}, and NdCoNi are presented in the Supplemental Material~\cite{supplemental}.}
    \label{fig:refined_PSI}
    
    \vspace{0.3cm} 
    
    \includegraphics[width=\textwidth]{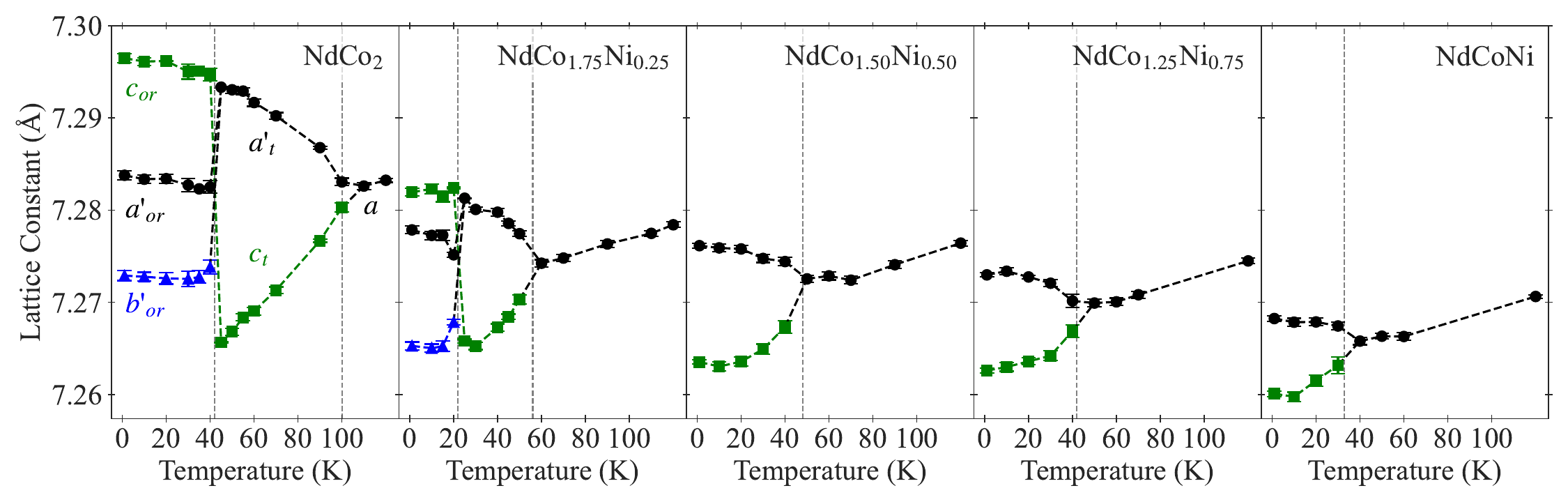}
    \caption{Unit cell axes obtained by Rietveld refinements of PND data at selected temperatures in the temperature interval 1.5 – 120~K. For the tetragonal and orthorhombic phases, the pseudo-cubic unit cell axes are shown for $a$ and $b$ ($a_{t}' =\sqrt{2}a_{t}$, $a_{or}' =\sqrt{2}a_{or}$, and $b_{or}' =\sqrt{2}b_{or}$). The error bars indicate the error of the refinements. Vertical gray lines indicate the temperatures of the phase transitions.}
    \label{fig:Lattice_parameters}
\end{figure*}

Rietveld refinements of the PND patterns for \ce{NdCo2} and \ce{NdCo_{1.75}Ni_{0.25}} are presented in Fig.~\ref{fig:refined_PSI}. No additional phases other than the main C15 phase were observed or included in the refinements. The refinements for the other studied compounds are presented in the Supplemental Material~\cite{supplemental}. The expected tetragonal and orthorhombic phase transitions induce small structural deviations from the cubic cell, resulting in minor but resolvable peak splittings. The tetragonal and orthorhombic distortions can be most efficiently identified from the (620) Bragg peak. This peak is marked with a star in Fig.~\ref{fig:refined_PSI} and shown in the insets. The evolution of the peak measured with SR-PXD from 80 to 300~K for \ce{NdCo2} is shown in the Supplemental Material~\cite{supplemental}. The small peak splitting shows that a careful analysis is required to assign the correct symmetry. 

All compounds exhibit the cubic structure (space group $Fd\bar{3}m$) at 120~K. This can, e.g., be seen by the absence of splitting of the (620) Bragg peak. In the cubic phase, the Nd atoms occupy the 8a sites, while the Co and Ni atoms occupy the 16d sites~\cite{xiao_crystal_2006}. The unit cell axis $a$ of \ce{NdCo2} at 120~K [Fig.~\ref{fig:Lattice_parameters}] agrees with previously reported values~\cite{xiao_crystal_2006}. The $a$ axis at this temperature shows a small and linear contraction with increasing Ni content (Supplemental Material~\cite{supplemental}), following a Vegard's law trend~\cite{vegard}. This is consistent with the larger lattice constants of \ce{NdCo2} (7.350~{\AA} at room temperature~\cite{gratz_isotropic_1994}) compared to \ce{NdNi2} (7.270~{\AA} at room temperature~\cite{osti_4121405}), and the similar behavior previously reported for \ce{PrNi_{2-x}Co_x} compounds~\cite{ermolenko_compositional_2019}.

Upon cooling, all compounds undergo a structural transition into a tetragonal phase (space group $I4_1/amd$), where the Nd atoms occupy the 4b sites, while the Co and Ni atoms occupy the 8c sites~\cite{xiao_crystal_2006}. The tetragonal transition occurs at 100~K for \ce{NdCo2}, consistent with previous findings~\cite{ouyang_magnetic_2005, xiao_crystal_2006, murtaza_magnetocaloric_2020}, and the transition temperature decreases with increasing Ni content. To avoid a jump in the plot of the unit cell axes resulting from the transitions from a face-centered cubic unit cell to a body-centered tetragonal unit cell, the pseudo-cubic lattice constants $a'_t$ of $a_t$ are presented in Fig.~\ref{fig:Lattice_parameters}, where $a'_{t} =\sqrt{2}a_{t}$. The difference between $a'_{t}$ and $c_{t}$ increases with decreasing temperature and decreases with increasing Ni substitution [Fig.~\ref{fig:Lattice_parameters}]. The tetragonal splitting can be observed from the (620) Bragg peak in Fig.~\ref{fig:refined_PSI} at 45~K, which is clearly split into two distinct peaks for \ce{NdCo2}. However, for \ce{NdCo_{1.75}Ni_{0.25}}, only a small shoulder and no distinct splitting is visible.

Upon cooling to 1.5~K, \ce{NdCo_{1.50}Ni_{0.50}}, \ce{NdCo_{1.25}Ni_{0.75}}, and NdCoNi remain in the tetragonal symmetry, while \ce{NdCo2} and \ce{NdCo_{1.75}Ni_{0.25}} undergo another phase transition into an orthorhombic phase with space group $Imma$~\cite{xiao_crystal_2006}. For this phase, the Nd atoms occupy the 4e sites, while the Co/Ni site is split into the 4a and 4d sites~\cite{xiao_crystal_2006}, as also supported by Mössbauer studies~\cite{atzmony_crystal-field-induced_1976}. As for the tetragonal phase, the pseudo-cubic $a'_{or} = \sqrt{2}a_{or}$ and $b'_{or} = \sqrt{2}b_{or}$ are shown in Fig.~\ref{fig:Lattice_parameters}. The differences between $a'_{or}$, $b'_{or}$, and $c_{or}$ are larger for \ce{NdCo2} than \ce{NdCo_{1.75}Ni_{0.25}} [Fig.~\ref{fig:Lattice_parameters}]. The (620) Bragg peak [Fig.~\ref{fig:refined_PSI}] is split into two peaks for \ce{NdCo2}, while for \ce{NdCo_{1.75}Ni_{0.25}} the orthorhombic distortion is only visible in the peak asymmetry and width.

\subsection{Magnetic structures from neutron diffraction}


PND was used to study the magnetic properties across the phase transitions. The cubic phases are paramagnetic, while the tetragonal and orthorhombic phases display long-range ferromagnetic order, described by the propagation vector $k = (0,0,0)$. All the magnetic scattering is thus located on nuclear Bragg peaks. This can, e.g., be seen at the low-$Q$ (111) peak at approximately 1.5~{\AA}$^{-1}$, which shows a magnetic contribution and increasing intensity with decreasing temperature across all compounds. However, the structural deformation is too small to use low-$Q$ peaks to distinguish the direction of the magnetic moments. Thus, we used the high-$Q$ (620) Bragg peak at approximately 5.45~{\AA}$^{-1}$ for this purpose, although the peak exhibits a moderate magnetic contribution due to its high $Q$ value~\cite{neutronref2}.

Since the thermal displacement parameter $B_{iso}$ exhibits a similar $Q$ dependence, accurate determination of $B_{iso}$ is essential to obtain a reliable magnetic moment from the refinement of the PND data. However, the decrease of the magnetic form factor is faster than the decrease of $B_{iso}$ as a function of $Q$. Therefore, short wavelength PND measurements ($\lambda$ = 1.154~{\AA}), providing a higher $Q$ range, were used to refine $B_{iso}$ at 1.5~K. This $B_{iso}$ was locked when refining the magnetic moment $M_{Nd}$ and $M_{Co/Ni}$. At 120~K, all compounds were paramagnetic, and $B_{iso}$ was refined, but not the magnetic moment. For intermediate temperatures, extrapolated $B_{iso}$ values were used to refine the magnetic moment. The temperature-dependence of $B_{iso}$ for all compounds is shown in the Supplemental Material~\cite{supplemental}.

Figure~\ref{fig:Magnetic_cells} shows the cubic, tetragonal, and orthorhombic unit cells for the \ce{NdCo_{2-x}Ni_x} compounds, with corresponding magnetic moment vectors. The cubic cell exhibits no magnetic moment. The $\Gamma_3^+$ irreducible representation describes the tetragonal deformation of the cubic cell, while the magnetic order follows $m\Gamma_4^+$, giving a magnetic space group $I4_1/am'd'$ [No. 141.557 Belov–Neronova–Smirnova (BNS)] in the tetragonal phase~\cite{talanov_magnetic_2023}. Our refinements show that the easy magnetization direction in this phase is along the [001] or $c$-direction, as reported previously based on PND~\cite{xiao_crystal_2006} and Mössbauer~\cite{atzmony_crystal-field-induced_1976} data. Magnetic moments are present on the Nd and the Co/Ni sites.

\begin{figure}[htbp]
  \centering
  {\includegraphics[height = 3.6cm]{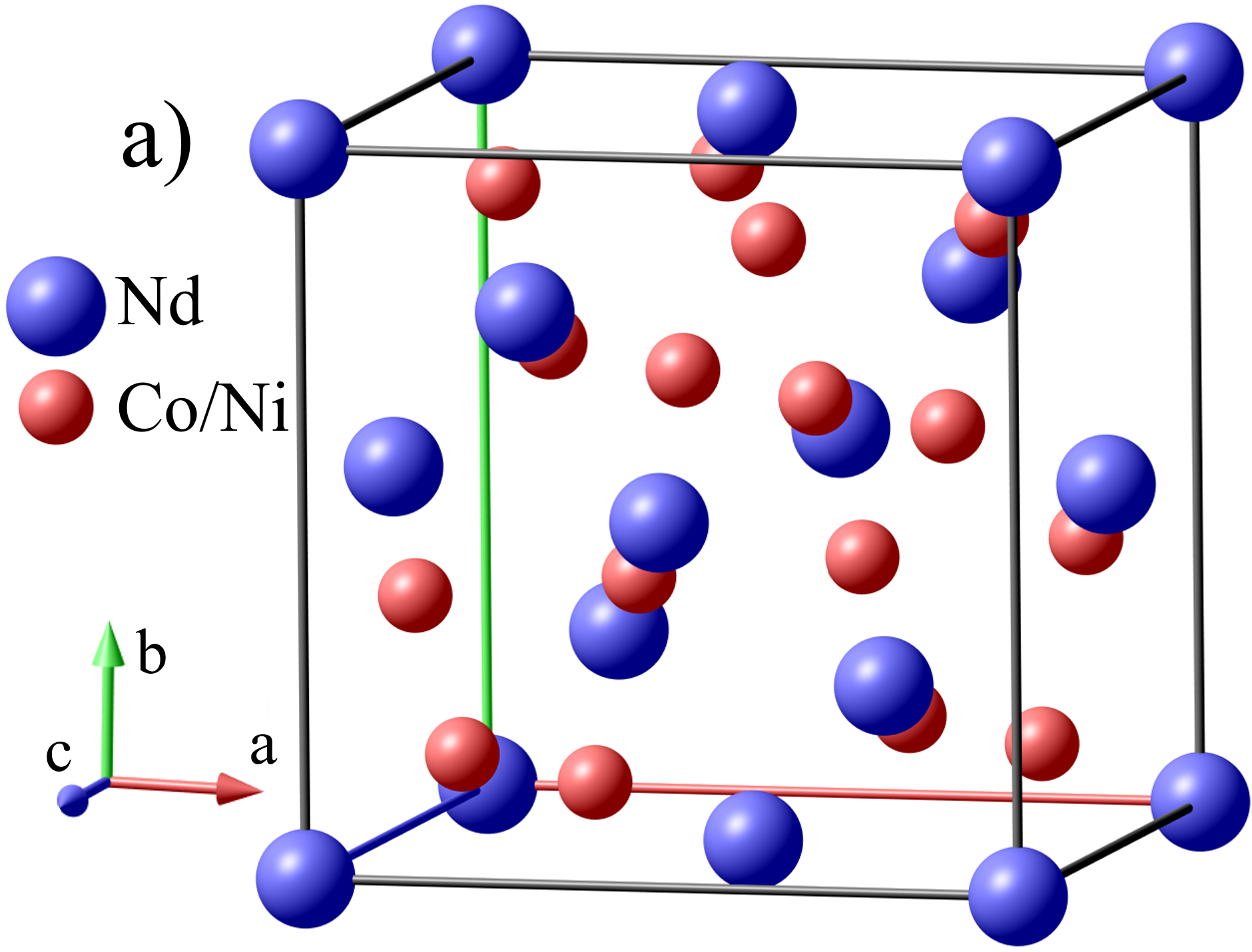} \label{fig:cubic_cell}} \\
  {\includegraphics[height = 3.6cm]{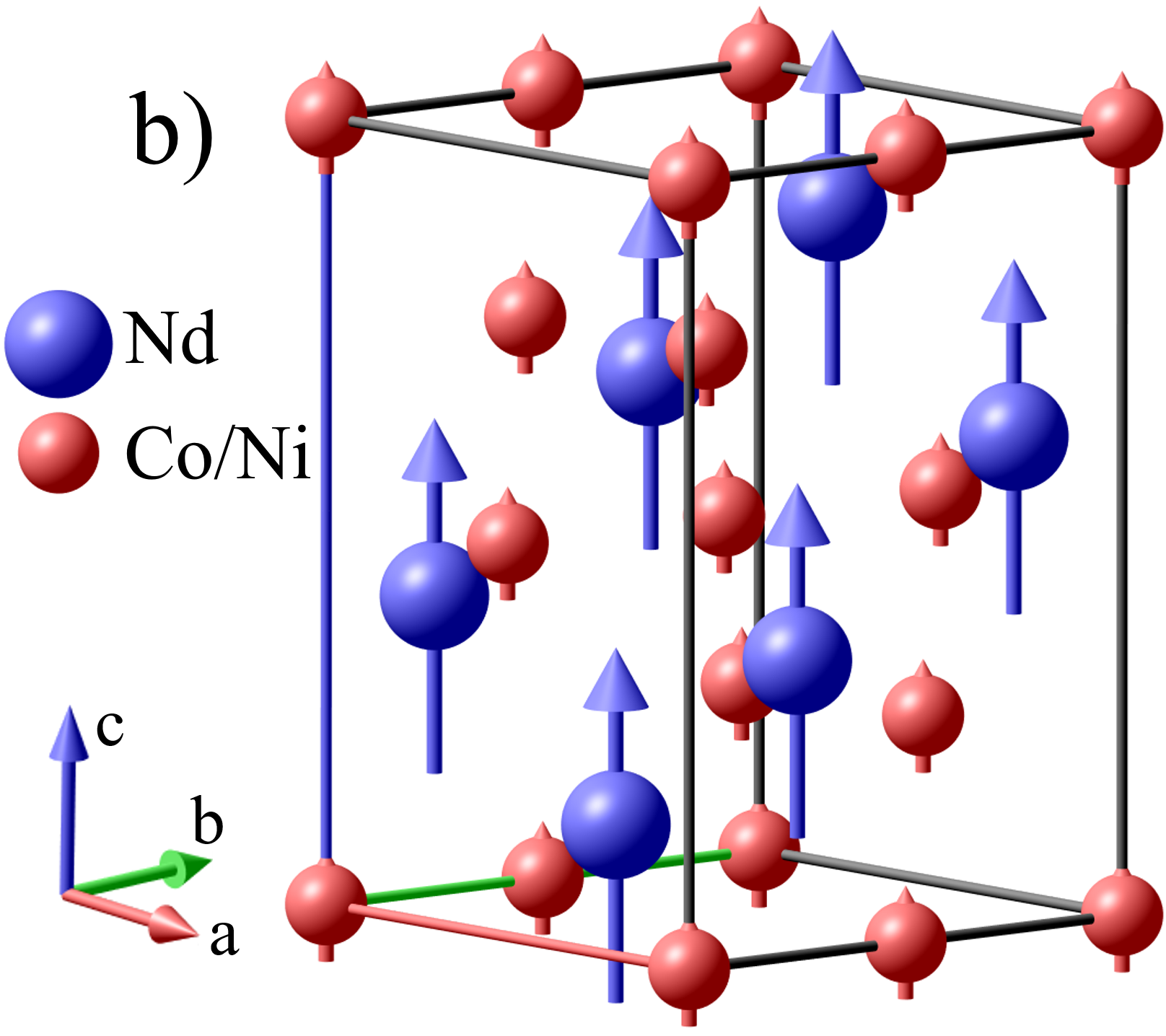} \label{fig:tetra_cell}}
  {\includegraphics[height = 3.6cm]{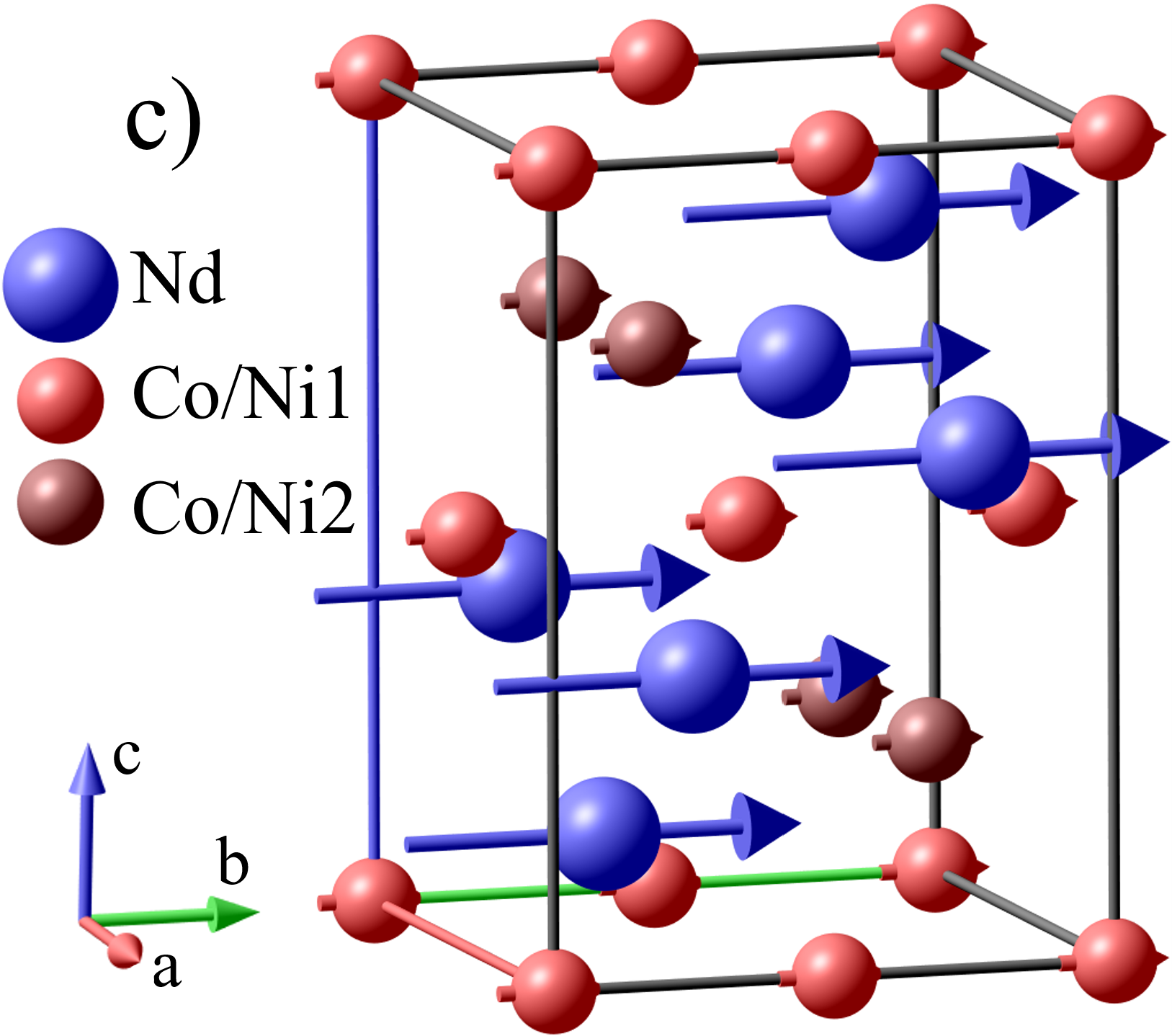} \label{fig:ortho_cell}} \\
  \caption{(a) Paramagnetic cubic unit cell, (b) tetragonal unit cell with magnetic moments aligned along the $c$ axis, and (c) orthorhombic unit cell with magnetic moments aligned along the $b$ axis of the \ce{NdCo_{2-x}Ni_{x}} compounds. The figures are created using Mag2Pol~\cite{mag2pol}.}
  \label{fig:Magnetic_cells}
\end{figure}

Upon further symmetry reduction to orthorhombic symmetry, the $\Gamma_5^+$ irreducible representation describes the structural distortion, and the magnetic symmetry is equivalently described by $m\Gamma_4^+$ or $m\Gamma_5^+$, both corresponding to $Imm'a'$ (BNS: 74.559)~\cite{talanov_magnetic_2023}. The magnetic moments are then reoriented to the $ab$ plane, marking the tetragonal-to-orthorhombic transition as a spin-reorientation transition. The origin of the spin rotation is based on changes in the crystal electric field~\cite{atzmony_crystal-field-induced_1976, dublon_crystal_1976}. The $Imm'a'$ space group does not distinguish between magnetic moments along the two short axes, as this corresponds to a positive or negative sign for the strain mode $\Gamma_5^+$. This is supported by Mössbauer measurements, where the easy magnetization direction in the orthorhombic phase of \ce{NdCo2} is determined to be in the [110] direction of the cubic unit cell, i.e., the $ab$ plane of the orthorhombic cell~\cite{atzmony_crystal-field-induced_1976}. 

\begin{figure}[htbp]
\centering
\includegraphics[width = 0.45\textwidth]{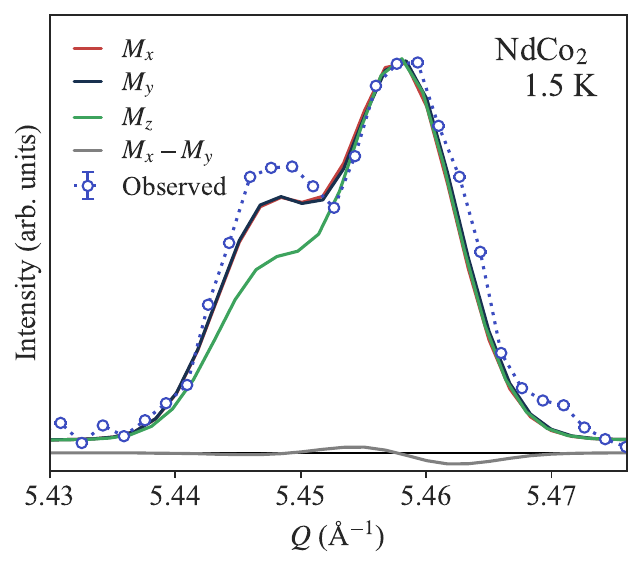}
\caption{Simulated PND patterns of the (620) Bragg peak using Mag2Pol with magnetic moments oriented in the $M_\text{x}$, $M_\text{y}$, and $M_\text{z}$ directions, respectively, compared to experimental data of the \ce{NdCo2} compound measured at 1.5~K. The difference between $M_\text{x}$ and $M_\text{y}$ is also shown.}
\label{fig:Simulated_patterns}
\end{figure}

A previous study on the refinement of PND measurements of \ce{NdCo2} has concluded that only the magnetic moments refined in the $b$-direction yield stable results~\cite{xiao_crystal_2006}. Figure~\ref{fig:Simulated_patterns} shows the differences between the refinement results in the present work when the magnetic moments are oriented along the $a$-, $b$-, and $c$-direction. The result for the $c$-direction is clearly different from the $a$- and $b$-directions, which are almost identical. Table~\ref{tab:direction} compares the magnetic moments and reliability factors for refinements of magnetic moments oriented along the $a$- and $b$-directions in the orthorhombic structure in the present work. 

\begin{table}[htbp]
\centering 
\caption{PND refinements of the magnetic moments oriented along the $a$ and $b$ unit cell axes in the orthorhombic phase. $M_\text{tot}$ is the magnitude of the total magnetic moment, while $\chi^2$ and $R_\text{wp}$ provide the reliability factors of the refinements.}
\begin{ruledtabular}
\begin{tabular}{l l c c}
Sample & Parameter & $a$ axis & $b$ axis \\
\hline \\[-2mm]
\ce{NdCo2}                & $M_\text{tot}$ ($\mu_\text{B}$) & 4.37(8) & 4.24(8) \\
                          & $\chi^2$ & 1.07 & 1.08 \\
                          & $R_\text{wp}$~(\%) & 3.37 & 3.40  \\
\ce{NdCo_{1.75}Ni_{0.25}} & $M_\text{tot}$ ($\mu_\text{B}$) & 3.77(9) & 4.05(5) \\
                          & $\chi^2$ & 1.16 & 1.16 \\
                          & $R_\text{wp}$~(\%) & 7.49 & 7.48 \\ 
\end{tabular}
\end{ruledtabular}
\label{tab:direction}
\end{table}

Based on our data and stable refinements, we cannot unambiguously determine whether the moments lie along $a$ or $b$ in the orthorhombic phase, as both solutions exhibit reasonable magnetic moments and errors. However, our refinements exclude the possibility of magnetic moments orienting along the $c$-direction in the orthorhombic phase, confirming a spin-reorientation transition from $M\parallel c$ to $M\perp c$ across the tetragonal-to-orthorhombic phase transition, as previously reported~\cite{dublon_crystal_1976}. Thus, the magnetic moments were oriented along the $b$-direction to be consistent with previous reports~\cite{xiao_crystal_2006}.

\begin{figure*}[htbp]
    \centering
    \includegraphics[width = \textwidth]{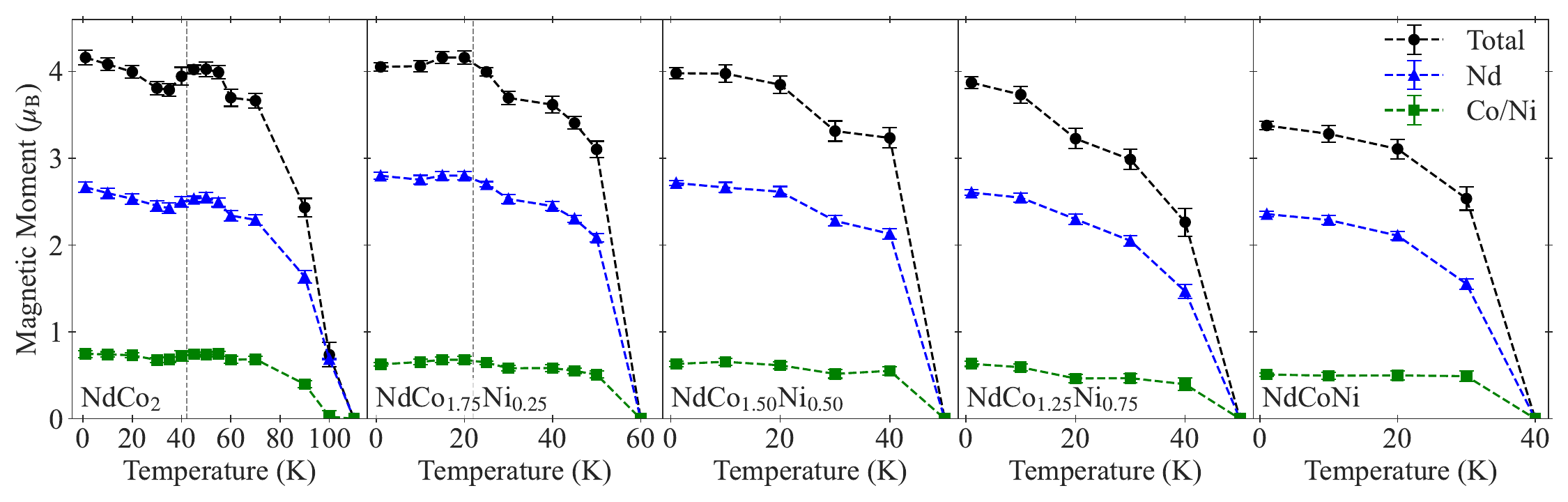}
    \caption{Atomic magnetic moments ($\mu_\text{B}$/atom) and total magnetic moment for the compounds ($\mu_\text{B}$/f.u.) as a function of temperature, derived from refinements of the PND data. Since Co and Ni occupy the same site, one cannot differentiate between them using PND. Therefore, the Co/Ni curves show an average between these two for each position. The total magnetic moment is calculated as the sum of the Nd and two times the Co/Ni magnetic moment. The vertical gray dotted lines mark $T_\text{SR}$ for \ce{NdCo2} and \ce{NdCo_{1.75}Ni_{0.25}}.}
    \label{fig:Magnetic_Moment}
\end{figure*}

Figure~\ref{fig:Magnetic_Moment} shows the temperature dependence of the atomic magnetic moments of the compounds, obtained from refinements of the PND data. The refinements were performed with magnetic moments oriented along the $c$ axis in the tetragonal cells and along the $b$ axis in the orthorhombic cells, as discussed above. All compounds are paramagnetic above the Curie temperature $T_\text{C}$, with increasing magnetic moments below this temperature. The magnetic moment of Nd is larger than that of Co/Ni for all compounds. Both the magnetic moments of Co/Ni and Nd decrease with increasing Ni substitution. There is a slight change in the magnitude of the magnetic moments of \ce{NdCo2} around $T_\text{SR}$.

The magnetic moments for \ce{NdCo2} are in line with those previously determined using PND~\cite{xiao_crystal_2006, ouyang_magnetic_2005}. The magnetic moment of Nd in \ce{NdCo2} at 1.5~K is 2.67~$\mu_\text{B}$, and as expected, somewhat lower than the theoretical value of $M_{cal} = g_\text{J} J = 3.27~\mu_\text{B}$. The magnetic moment of Co in \ce{NdCo2} at 1.5~K is 2$\cdot$0.75~$\mu_\text{B}/f.u.$ and the Ni/Co moment in NdCoNi is 2$\cdot$0.51~$\mu_\text{B}/f.u$. Assuming a constant Co moment across the series gives 0.27~$\mu_\text{B}/f.u.$ for the moment of Ni in NdCoNi.

\subsection{Bulk magnetic properties}

\begin{figure*}[htbp]
    \centering
    \includegraphics[width = 0.49\textwidth]{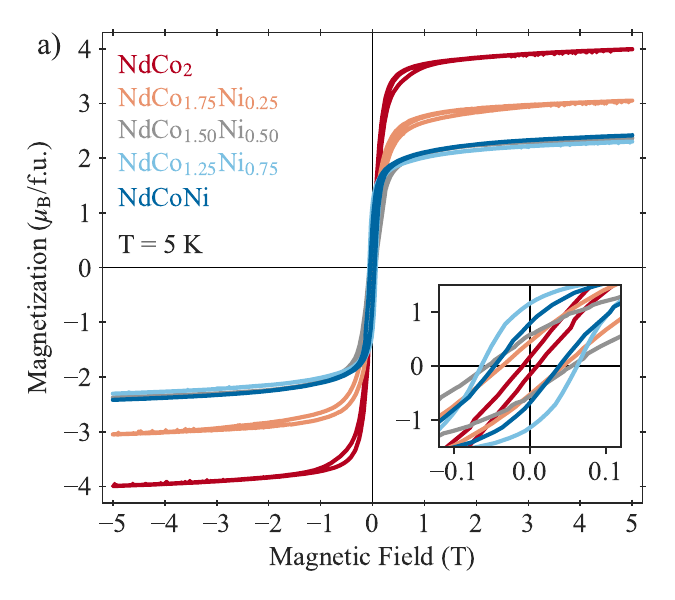}
    \includegraphics[width = 0.49\textwidth]{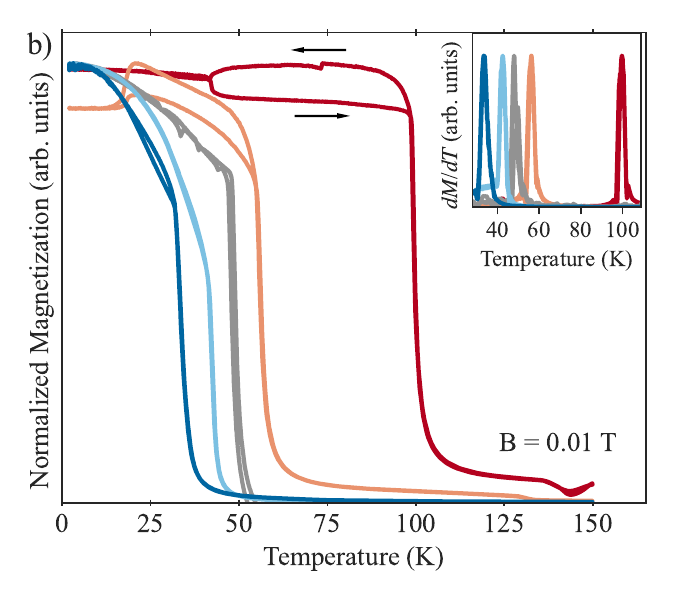}
    \caption{Bulk magnetization for the five selected compositions of the \ce{NdCo_{2-x}Ni_{x}} system. (a) Field-dependent magnetization ($-5$~T~$\leq~B~\leq~5$~T) measured at 5~K, where the inset displays the magnetization around the origin. (b) Temperature-dependent magnetization ($5~$K$~ \leq~T~\leq~150$~K) at 0.01~T, where the inset shows the derivative of the magnetization as a function of temperature. The maximum of the derivative was used here to determine $T_\text{C}$. The arrows indicate the magnetization data corresponding to heating and cooling.}
    \label{fig:Magnetic}
\end{figure*}

Figure~\ref{fig:Magnetic} shows bulk magnetization results. Values for the saturation magnetization $M_\text{S}$, remanent magnetization $M_\text{R}$, and coercivity $H_\text{C}$ determined from these measurements are given in the Supplemental Material~\cite{supplemental}, along with temperature-dependent magnetization measurements performed at fields of 1 and 5~T, respectively. All compounds exhibit a small magnetic hysteresis at low fields and magnetization. \ce{NdCo2} exhibits the smallest hysteresis, and \ce{NdCo_{1.25}Ni_{0.25}} the largest.

The bulk $M_\text{S}$ of \ce{NdCo2} (3.99~$\mu_\text{B}$) is slightly lower than the total magnetic moment $M_\text{tot}$ derived from PND refinements (4.24~$\mu_\text{B}$), which can be attributed to a lack of saturation at the applied field of 5~T. The bulk $M_\text{S}$ decreases drastically upon substitution of only \ce{Ni_{0.25}}, from 3.99 to 3.05~$\mu_\text{B}$. There is a smaller but still significant decrease upon substitution to \ce{Ni_{0.50}} (to 2.37~$\mu_\text{B}$), while subsequent Ni substitution has a negligible effect on $M_\text{S}$. The reduction in $M_\text{S}$ upon Ni substitution of Co is significantly stronger than expected based on their difference in magnetic moments according to Hund's rules~\cite{blundell_magnetism_2001}, and the PND results [Fig.~\ref{fig:Magnetic_Moment}]. However, a similar strong change in magnetic moment has been observed in bulk magnetization measurements of \ce{PrCo_{2-x}Ni_x} as a function of $x$~\cite{ermolenko_compositional_2019}. It is known that Co couples ferromagnetically with light rare-earth elements in cubic Laves compounds~\cite{farrell_magnetic_1966}. The reduction of $M_\text{S}$ in \ce{PrCo_{2-x}Ni_x} has been attributed to a strong decrease in the exchange interactions from \ce{PrCo2} to \ce{PrCo_{1.8}Ni_{0.2}} as a result of the Co atoms being replaced by the more weakly magnetic Ni as the nearest neighbors to Pr~\cite{ermolenko_compositional_2019}. A similar explanation could apply to the trends observed in this work, but additional studies are needed to confirm this.

The Curie temperature was determined from the maximum of the derivative of the temperature-dependent magnetization, as shown in Fig.~\ref{fig:Magnetic}(b) inset, using an applied magnetic field of 0.01~T to minimize field-induced shifts. The $T_\text{C}$ of \ce{NdCo2} (100~K) and NdCoNi (34~K) are consistent with previous reports~\cite{xiao_crystal_2006,lunde_machine_2025}, while the intermediate stoichiometries exhibit values between those of the two end members (56, 48, and 42~K, respectively). A similar trend is observed in \ce{PrNi_{2-x}Co_x}~\cite{ermolenko_compositional_2019}, where $T_\text{C}$ decreases with increasing Ni substitution. This can be attributed to weaker exchange interactions as the Ni content, which strongly influences $T_\text{C}$ in $R$\ce{Co2} compounds, where $R$ is a rare-earth element~\cite{murtaza_magnetocaloric_2020}.

A small thermal hysteresis can be observed in the tetragonal phase [Fig.~\ref{fig:Magnetic}(b)], the cubic phase is paramagnetic, and the orthorhombic phase exhibits no visible hysteresis. Thus, the spin reorientation associated with the tetragonal-to-orthorhombic phase transition can be observed in the bulk magnetization data at the low-temperature end of the hysteresis for \ce{NdCo2} (42~K) and \ce{NdCo_{1.75}Ni_{0.25}} (22~K). This $T_\text{SR}$ can be observed as a small bump at the temperature-dependent curves measured at 1~T, whereas no transition is visible at 5~T, as the high external field suppresses it (see Supplemental Material~\cite{supplemental}).

\subsection{Magnetocaloric effect}

\begin{figure}[htbp]
  \centering
  \includegraphics[width = 0.49\textwidth]{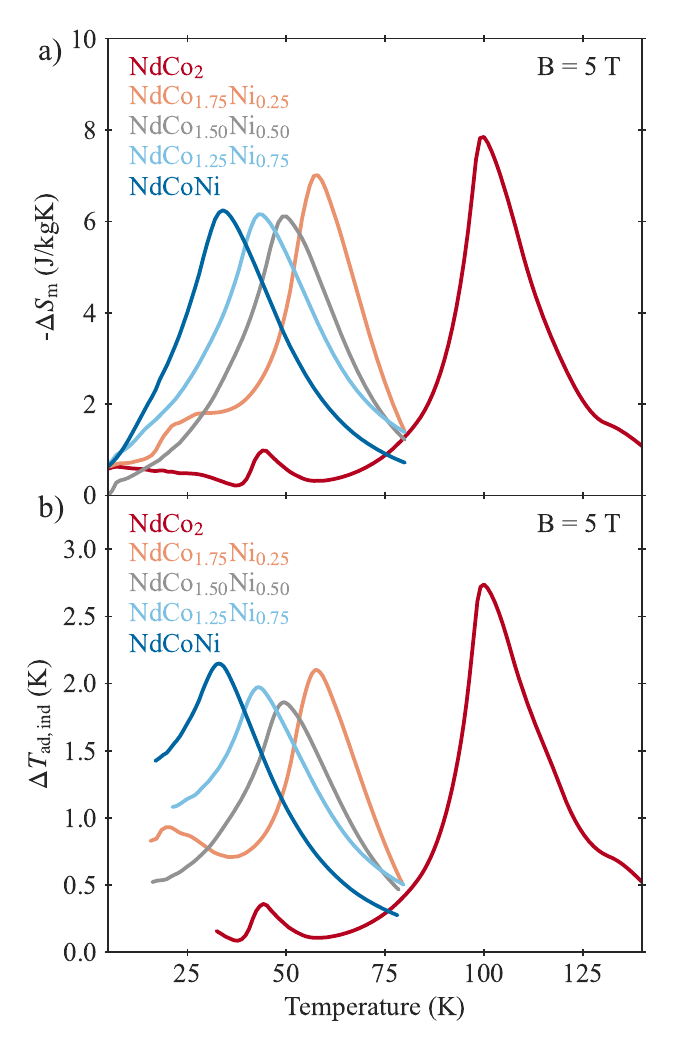} \label{fig:dS_dT} \\
  \vspace{-0.7cm} 
  \caption{(a) $\Delta S_\text{m}$ and (b) $\Delta T_\text{ad,ind}$ results obtained at 5~T from S-T diagrams based on $C_\text{P}$ measurements.}
  \label{fig:MCE}
\end{figure}

\begin{figure}[htbp]
  \centering
  \includegraphics[width = 0.49\textwidth]{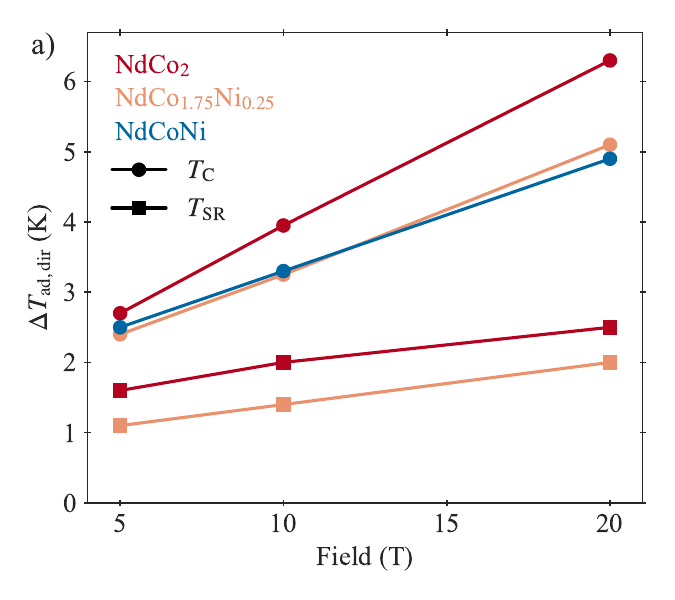} \label{fig:Pulsed_Fields}\\
  \vspace{-0.8cm} 
  \includegraphics[width = 0.49\textwidth]{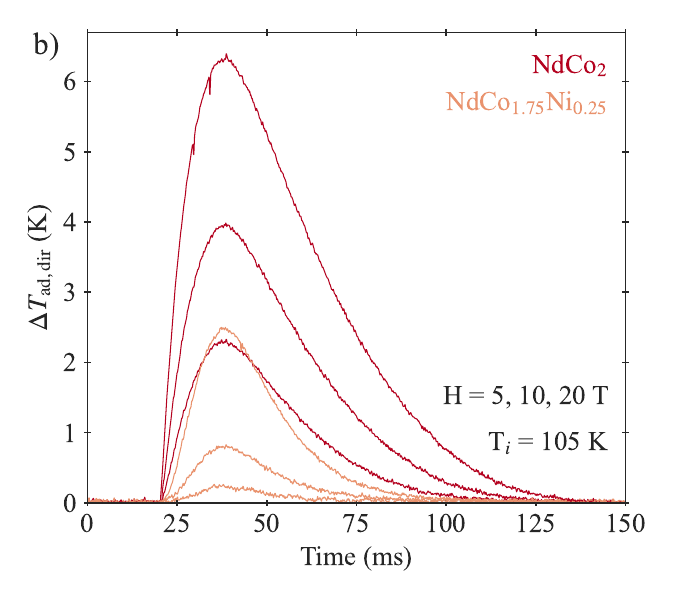} \label{fig:Pulse_time} \\
  \vspace{-0.7cm} 
  \caption{(a) Pulsed-field measurements of $\Delta T_\text{ad,dir}$ around $T_\text{C}$ of \ce{NdCo2} (100~K), \ce{NdCo_{1.75}Ni_{0.25}} (56~K), and NdCoNi (34~K), and $T_\text{SR}$ of \ce{NdCo2} (42~K) and \ce{NdCo_{1.75}Ni_{0.25}} (22~K). (b) Time-dependent $\Delta T_\text{ad,dir}$ during the magnetic field pulse applied to the \ce{NdCo2} and \ce{NdCo_{1.75}Ni_{0.25}} compounds at 105~K.}
  \label{fig:Pulsed}
\end{figure}

To determine the order of the phase transition, an $n$ exponent analysis was performed using the $\Delta S_\text{m}$ curves obtained from the $M(H)$ data. As reported in~\cite{law_quantitative_2018}, a local $n$ exponent greater than 2 signifies a first-order magnetic transition. This analysis confirms that the transitions at $T_\text{SR}$ of \ce{NdCo2} and \ce{NdCo_{1.75}Ni_{0.25}} (at 42 and 22~K, respectively) are first-order phase transitions, while the transitions of all compounds at $T_\text{C}$ are second-order phase transitions.

The MCE was characterized by $\Delta S_\text{m}$ and $\Delta T_\text{ad,ind}$ for all samples. Additionally, $\Delta T_\text{ad,dir}$ was determined by pulsed-field measurements for \ce{NdCo2}, \ce{NdCo_{1.75}Ni_{0.25}}, and NdCoNi. $\Delta S_\text{m}$ and $\Delta T_\text{ad,ind}$ at 5~T extracted from $S$-$T$ diagrams are shown in Fig.~\ref{fig:MCE}, while the $\Delta T_\text{ad,dir}$ at 5, 10, and 20~T are shown in Fig.~\ref{fig:Pulsed}. A comparison of the results obtained using the different methods is presented in the Supplemental Material~\cite{supplemental}. The agreement between $\Delta T_\text{ad,dir}$ and $\Delta T_\text{ad,ind}$ is satisfactory, with the direct method yielding slightly higher values. $\Delta T_\text{ad,dir}$ should be the more reliable method in particular for the first-order phase transitions, i.e., $T_\text{SR}$~\cite{politova_magnetism_2024}. This is particularly evident for the \ce{NdCo2} sample in the present work.

A strong MCE is associated with $T_\text{C}$ of all compounds, and with $T_\text{SR}$ of \ce{NdCo2} and \ce{NdCo_{1.75}Ni_{0.25}}. The MCEs at $T_\text{C}$ exceed those observed at $T_\text{SR}$. The Ni-substituted compounds exhibit MCEs that span most of the temperature region needed for magnetocaloric hydrogen liquefaction (20 to 77~K), whereas the MCE of \ce{NdCo2} appears at higher temperatures. A slight decrease in $\Delta S_\text{m}$ is observed with increasing Ni substitution, but $\Delta T_\text{ad}$ does not follow this trend, except for \ce{NdCo2} displaying the highest value. \ce{NdCo2} has previously been reported to exhibit, at $T_\text{C}$, $\Delta S_\text{m}$ of 7.3~J/kgK and a $\Delta T_\text{ad}$ of 3.1~K at 5~T~\cite{murtaza_magnetocaloric_2020}. In comparison, the $\Delta S_\text{m}$ we obtained in the present study (9.0~J/kgK) is higher, while the $\Delta T_\text{ad}$ is comparable when determined from $M(H)$ measurements (3.1~K). We have previously studied the MCE of NdCoNi~\cite{lunde_machine_2025}, yielding slightly higher values for $\Delta S_\text{m}$ and $\Delta T_\text{ad,ind}$ compared to the present work using the $M(H)$ method but slightly lower values using the $C_\text{P}$ method, respectively.

Figure~\ref{fig:Pulsed}(a) shows the evolution of $\Delta T_\text{ad,dir}$ at $T_\text{SR}$ (\ce{NdCo2} and \ce{NdCo_{1.75}Ni_{0.25}}) and $T_\text{C}$ (\ce{NdCo2}, \ce{NdCo_{1.75}Ni_{0.25}} and \ce{NdCoNi}) for fields up to 20~T. As expected, $\Delta T_\text{ad,dir}$ is lower than for heavy rare-earth-based compounds measured using similar techniques, including $R$\ce{Co2} Laves compounds where heavy rare-earth elements $R$ were used~\cite{bykov_magnetocaloric_2024, politova_magnetism_2024, dragland_relation_2025}. $\Delta T_\text{ad,dir}$ returned reversibly to 0~K by 150~ms for all applied field [Fig.~\ref{fig:Pulsed}(b)]. The MCE at $T_\text{C}$ increases more steeply than at $T_\text{SR}$ with increasing field strength. This difference was quantified by calculating $[\Delta T_\text{ad,dir}$($T_\text{SR}$)/$\Delta T_\text{ad,dir}$($T_\text{C}$)]$\cdot 100~\%$. At 5~T, this ratio is 58 and 46~\% for \ce{NdCo2} and \ce{NdCo_{1.75}Ni_{0.25}}, respectively, while at 20~T the values are 45 and 39~\%, respectively. Additionally, there is a difference between the field dependence of $\Delta T_\text{ad,dir}$ of the different compounds. Considering $T_\text{C}$, NdCoNi exhibits a slightly higher $\Delta T_\text{ad,dir}$ than \ce{NdCo_{1.75}Ni_{0.25}} at 5~T, while at 20~T the $\Delta T_\text{ad,dir}$ of \ce{NdCo_{1.75}Ni_{0.25}} is higher. The $\Delta T_\text{ad,dir}$ at $T_\text{C}$ of \ce{NdCo2} exhibits an even stronger dependence on applied field.

\section{Discussion} 

Ni substitution of Co in \ce{NdCo2} reduces the tetragonal and orthorhombic distortions [Fig.~\ref{fig:refined_PSI} and Fig.~\ref{fig:Lattice_parameters}]. This observation can be attributed to differences in the atomic radii of Co and Ni. The same trend is seen for \ce{NdCo_{2-x}Fe_x}, where the larger Fe atoms increase the difference between $a'_{t}$ and $c_{t}$ in the tetragonal unit cell, from 0.032~{\AA} for \ce{NdCo2} to 0.048~{\AA} for \ce{NdCo_{1.4}Fe_{0.6}} ~\cite{murtaza_magnetocaloric_2020}.

\begin{figure}[htbp]
\centering
\includegraphics[width = 0.5\textwidth]{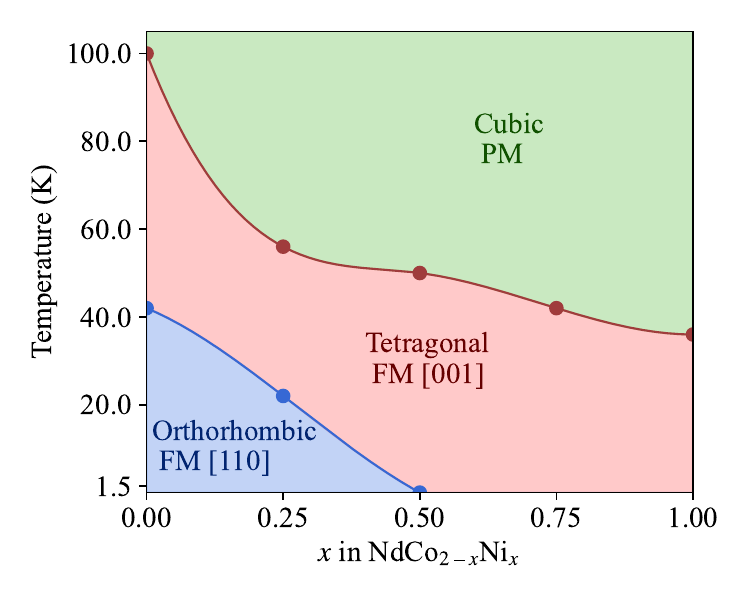}
\vspace{-0.7cm}
\caption{Phase diagram over the paramagnetic (PM) cubic, ferromagnetic (FM) tetragonal, and FM orthorhombic phases, with the directions of the magnetic moments. The solid circles indicate measured transition temperatures, while the lines are interpolated between these.}
\label{fig:Phase_diagram}
\end{figure}

Furthermore, the Ni substitution shifts both the tetragonal and the orthorhombic phase transitions towards lower temperatures. Figure~\ref{fig:Phase_diagram} shows the phase diagram of the \ce{NdCo_{2-x}Ni_x} system, depicting the various crystalline phases present over the temperature range from 1.5 to 105~K. Since both transitions are coupled magnetostructural, their shifts can be attributed either to a shift in the exchange constant driving $T_\text{C}$~\cite{liu_matter_2024}, or the difference in atomic radii of Ni and Co driving the structural distortion. A similar shift in the transition temperature was seen in \ce{NdCo_{2-x}Fe_x}, attributed to the Fe substitution increasing $T_\text{C}$, and thus the tetragonal phase transition temperature~\cite{murtaza_magnetocaloric_2020}.

The direction of the magnetic moments in the orthorhombic phase could not be determined unambiguously between the $a$- and $b$-directions from our PND data. There are minor differences between the simulated patterns of moments oriented along these two directions for the high $Q$ (620) Bragg peak of \ce{NdCo2} [Fig.~\ref{fig:Simulated_patterns}], where the peak splitting is observed for the orthorhombic phase. Unambiguous determination of the direction of the magnetic moments could be carried out with single-crystal neutron diffraction on an untwinned crystal. For compounds with Ni substitution, the peak splitting decreases with increasing substitution, requiring even higher resolution to determine the direction of the magnetic moments. 

The $M_\text{tot}$ determined from PND refinements [Fig.~\ref{fig:Magnetic_Moment}] is similar to the $M_\text{S}$ determined from bulk magnetization measurements [Fig.~\ref{fig:Magnetic}(a)] for \ce{NdCo2}, 4.24 and 3.99~$\mu_\text{B}$, respectively. However, upon substitution of Ni, $M_\text{S}$ decreases more rapidly than $M_\text{tot}$. For NdCoNi, $M_\text{tot}$ is 3.38~$\mu_\text{B}$, while $M_\text{S}$ is 2.42~$\mu_\text{B}$. 
Part of the discrepancy might be attributed to the measurement temperature of $M_\text{S}$ (5~K) being closer to $T_\text{C}$ than the measurement temperature of $M_\text{tot}$ (1.5~K), in particular for compounds with high Ni substitution where $T_\text{C}$ is low. The field-dependent bulk magnetization [Fig.~\ref{fig:Magnetic}(a)] of all the compounds is not completely saturated at 5~T, but increases by approximately 0.05~$\mu_\text{B}$/T. As we have seen from x-ray magnetic circular dichroism experiments of NdCoNi, the increase can be attributed to van Vleck paramagnetism of Nd~\cite{lunde_electronic_2025, wallace1968magnetic}.

The $n$ exponent analysis applied to the $M(H)$ data agrees with most of the present published data that the phase transition at $T_\text{C}$ of \ce{NdCo2} is of the second-order while the transition at $T_\text{SR}$ is of the first-order (Supplemental Material~\cite{supplemental}). As the Ni substitution increases along the \ce{NdCo_{2-x}Ni_x} series, the $n$ exponent approaches 2 more closely, although the differences between the compositions are small.  The $n$ exponent does not overshoot above 2 for any of the compounds, which would indicate a first-order phase transition~\cite{law_quantitative_2018}.
$n$ decreases away from 2 in high fields, where the transition is suppressed~\cite{bykov_magnetocaloric_2024}.

The decrease in MCE from \ce{NdCo2} upon Ni substitution is attributed to the reduced magnetic moment. The highest MCE was observed at $T_\text{C}$, rendering the suppression of $T_\text{SR}$ at high Ni content irrelevant for applications. The compounds studied in this work cover the lower end of the temperature range required for magnetocaloric hydrogen liquefaction, while there is a gap at the higher end, near 77~K. The Ni substitution allows tuning of $T_\text{C}$, enabling the possibility of covering the entire magnetocaloric hydrogen liquefaction region from 20 to 77~K by fine-tuning the stoichiometry of \ce{NdCo_{2-x}Ni_x} with 0 $\leq$ x $\leq$ 0.25. 

\section{Conclusions} 

The structural and magnetic phase transitions, as well as the magnetocaloric effects, of five \ce{NdCo_{2-x}Ni_x} ($x=$ 0, 0.25, 0.50, 0.75, 1) cubic Laves compositions have been studied for possible magnetic refrigeration applications in a temperature range relevant for hydrogen liquefaction. 
All compounds display a phase transition from a paramagnetic cubic phase to a ferromagnetic tetragonal phase with magnetic moments oriented along the $c$-direction upon cooling. The Co-rich samples undergo a spin-reorientation transition associated with a further symmetry reduction to an orthorhombic space group. Both the tetragonal and orthorhombic distortions and transition temperatures decreased with increasing Ni substitution. Bulk magnetization measurements show a larger decrease than neutron diffraction measurements. 
The phase transition orders were verified using $n$ exponent analysis. The magnetocaloric effect decreased slightly with increasing Ni substitution due to a reduced magnetic moment. The adiabatic temperature change was measured directly at 5, 10, and 20~T and was observed to scale more strongly with the applied field at the Curie temperature than at the spin-reorientation temperature.

\begin{acknowledgments}
This work was financed by the Research Council of Norway under project number 336403. The authors gratefully acknowledge support from the Clean Hydrogen partnership and its members within the framework of the project HyLICAL (Grant No. 101101461). The authors also acknowledge NcNeutron, financed by the Research Council of Norway under project number 245942. The authors thank the staff at SNBL (BM01 and BM31) of the ESRF for their skilled assistance during the measurements. Part of this work was performed at the Swiss Spallation Neutron Source (SINQ), Paul Scherrer Institut (PSI), Villigen, Switzerland. We acknowledge support from Deutsche Forschungsgemeinschaft (DFG) through CRC/TRR 270 (Project-ID 405553726) and through W\"urzburg-Dresden Cluster of Excellence on Complexity and Topology in Quantum Matter---$ct.qmat$ (EXC 2147, Project No.\ 390858490) and the support of the HLD at HZDR, member of the European Magnetic Field Laboratory (EMFL).
\end{acknowledgments}

\section*{Data availability}

The data that support the findings of this article are openly available~\cite{PT4R3F_2025}.

\end{document}


\preprint{APS/123-QED}

\title{Supplemental material for "Influence of Ni substitution on the phase transitions and magnetocaloric effect of \ce{NdCo2} at cryogenic temperatures"}

\author{Vilde G. S. Lunde}\thanks{Contact author: vilde.lunde@ife.no}
\affiliation{Department for Hydrogen Technology, Institute for Energy Technology (IFE), PO Box 40, NO-2027, Kjeller, Norway}
\author{{\O}ystein S. Fjellv{\aa}g}
\affiliation{Department for Hydrogen Technology, Institute for Energy Technology (IFE), PO Box 40, NO-2027, Kjeller, Norway}
\author{Allan M. Döring}
\affiliation{Functional Materials, Institute of Materials Science, Technical University of Darmstadt, Darmstadt, 64287, Germany}
\author{Marc Stra{\ss}heim}
\affiliation{Dresden High Magnetic Field Laboratory (HLD-EMFL) and W\"urzburg-Dresden Cluster of Excellence ct.qmat, Helmholtz-Zentrum Dresden-Rossendorf, 01328 Dresden, Germany}
\affiliation{Institut für Festkörper- und Materialphysik, Technische Universität Dresden, 01069 Dresden, Germany}
\author{Vladimir Pomjakushin}
\affiliation{Paul Scherrer Institute (PSI), 5232, Villigen, Switzerland}
\author{Konstantin P. Skokov}
\affiliation{Functional Materials, Institute of Materials Science, Technical University of Darmstadt, Darmstadt, 64287, Germany}
\author{Oliver Gutfleisch}
\affiliation{Functional Materials, Institute of Materials Science, Technical University of Darmstadt, Darmstadt, 64287, Germany}
\author{Tino Gottschall}
\affiliation{Dresden High Magnetic Field Laboratory (HLD-EMFL) and W\"urzburg-Dresden Cluster of Excellence ct.qmat, Helmholtz-Zentrum Dresden-Rossendorf, 01328 Dresden, Germany}
\author{Joachim Wosnitza}
\affiliation{Dresden High Magnetic Field Laboratory (HLD-EMFL) and W\"urzburg-Dresden Cluster of Excellence ct.qmat, Helmholtz-Zentrum Dresden-Rossendorf, 01328 Dresden, Germany}
\affiliation{Institut für Festkörper- und Materialphysik, Technische Universität Dresden, 01069 Dresden, Germany}
\author{Anja O. Sj{\aa}stad}
\affiliation{Department of Chemistry and Centre for Material Science and Nanotechnology (SMN), University of Oslo, NO-0315, Norway}
\author{Bj{\o}rn C. Hauback}
\affiliation{Department for Hydrogen Technology, Institute for Energy Technology (IFE), PO Box 40, NO-2027, Kjeller, Norway}
\author{Christoph Frommen} \thanks{Contact author: christoph.frommen@ife.no}
\affiliation{Department for Hydrogen Technology, Institute for Energy Technology (IFE), PO Box 40, NO-2027, Kjeller, Norway}

\renewcommand{\thefigure}{A\arabic{figure}}
\setcounter{figure}{0}
\renewcommand{\thetable}{A\arabic{table}}
\setcounter{table}{0}
\renewcommand{\theequation}{A\arabic{equation}}
\setcounter{equation}{0}

\date{\today}


\begin{abstract}
The supplemental material contains scanning electron microscope-energy-dispersive x-ray (SEM-EDS) results [Section~\ref{sec:SEM}], synchrotron radiation powder x-ray diffraction (SR-PXD) data [Section~\ref{sec:XRD}], powder neutron diffraction (PND) data [Section~\ref{sec:PND}], as well as bulk magnetization and magnetocaloric effect data [Section~\ref{sec:PPMS}].
\end{abstract}

\maketitle

\section{SEM-EDS Analysis}
\label{sec:SEM}

\begin{figure*}[htbp]
    \centering
    \includegraphics[height = 6cm]{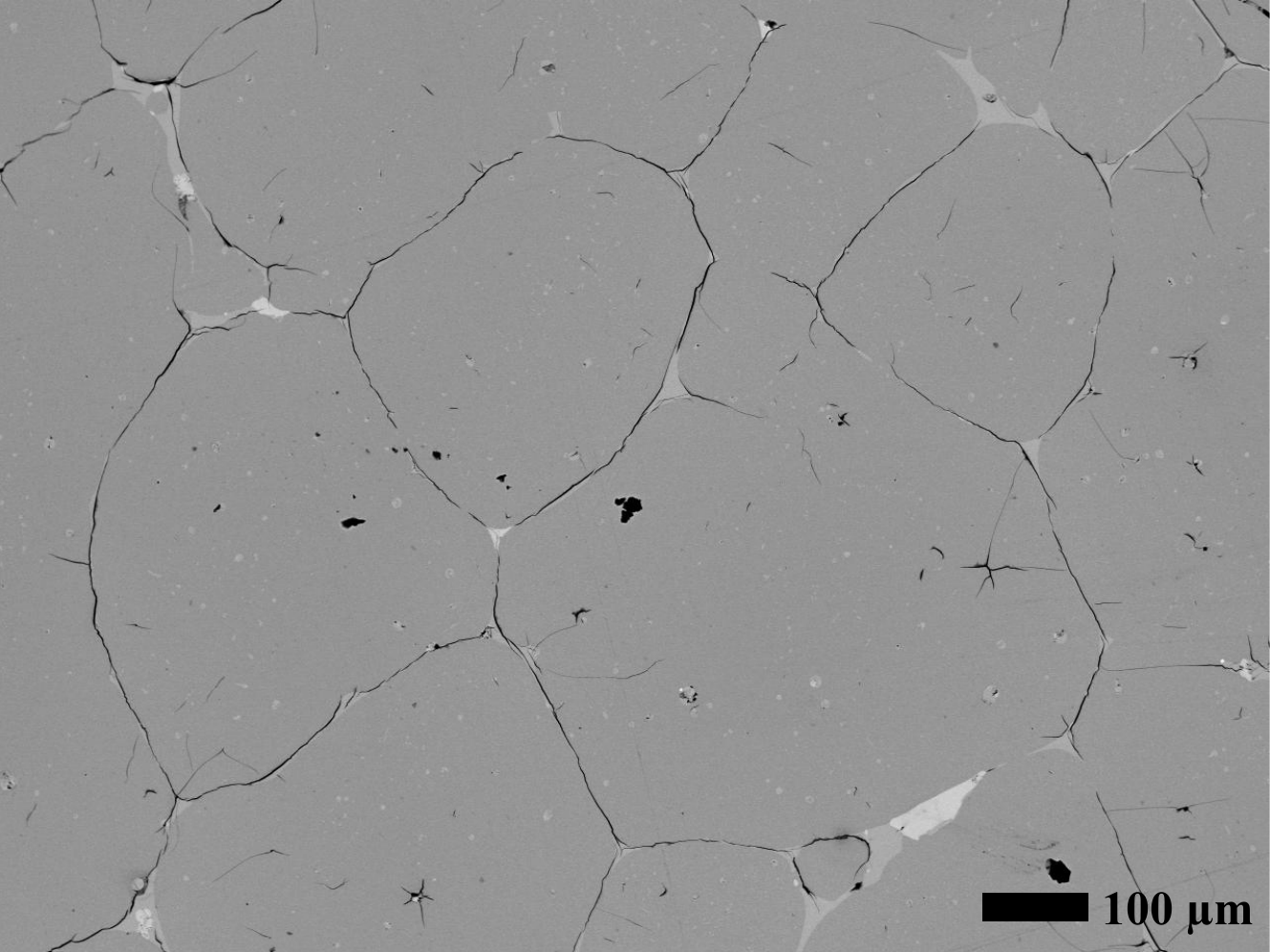}
    \includegraphics[height = 6cm]{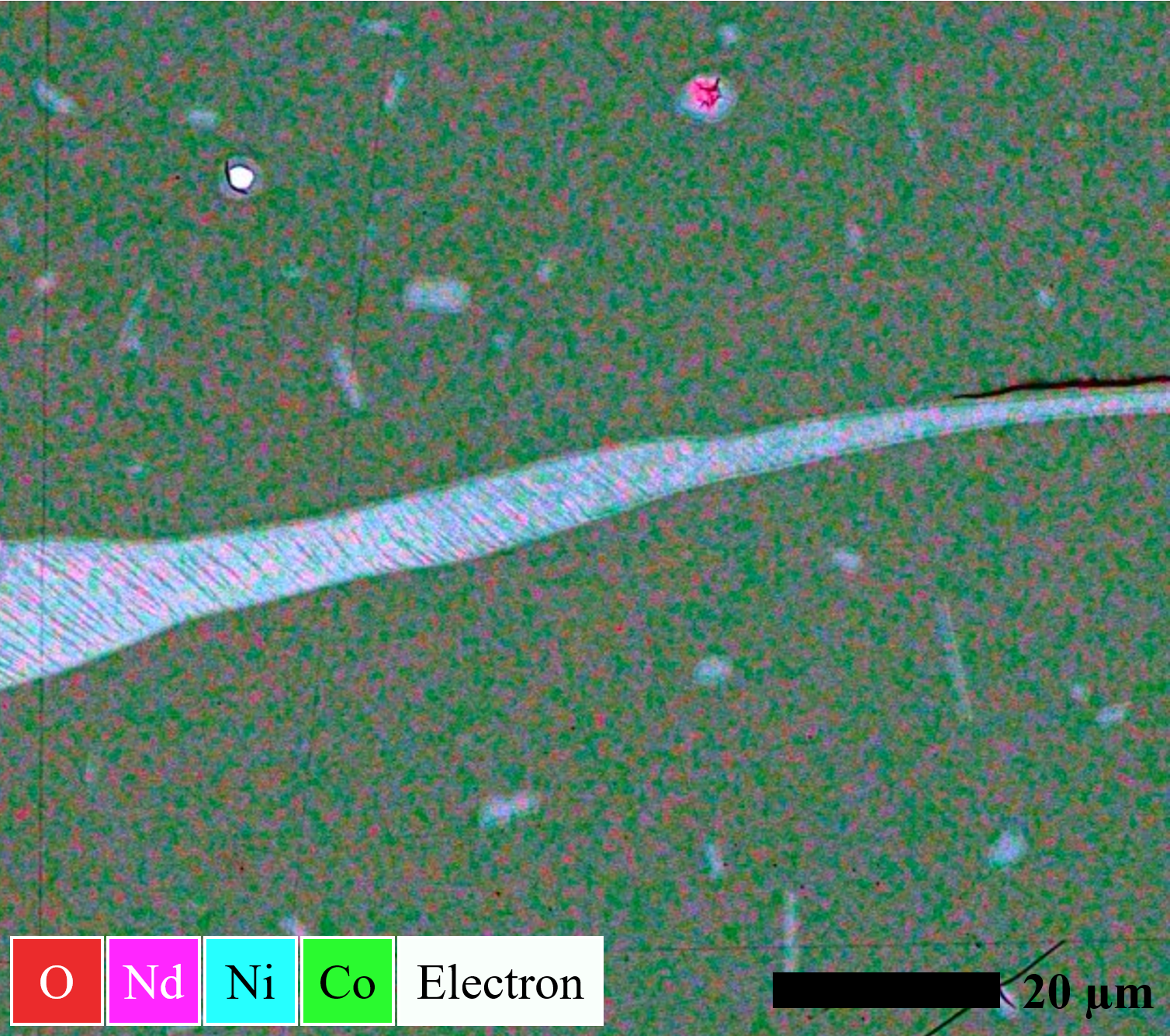}
    \caption{Representative back-scattered electron image (left) and EDS measurements (right) of the \ce{NdCo_{1.50}Ni_{0.50}} compound, illustrating the microstructure typical for the compounds in this study. The main phase is the \ce{AB2} Laves phase. There is a small oxygen contamination because the SEM-EDS sample preparation was performed in ambient air. The lighter areas are a \ce{NdNi} secondary phase, thus Co-deficient. Percentages of secondary phases for all analyzed compounds are tabulated in Table~\ref{tab:EDS}.}
    \label{fig:SEM}
\end{figure*}

\newpage

\begin{table}[htbp]
\centering 
\caption{Composition and quantity (area~\%) of secondary phases according to SEM-EDS analysis, determined as illustrated in Fig.~\ref{fig:SEM}. The remaining percentage corresponds to the primary \ce{AB2} Laves phase, where A is Nd and B is the stoichiometric transition metal composition of the given compound. The area percentage of the secondary phases is calculated from the average and standard deviation across three BSE images acquired at different positions within the same cross-section, all using the same scale.}
\begin{ruledtabular}
\begin{tabular}{l l l}
Primary \ce{AB_2} phase & Secondary phase & area~\% \\[1mm]
\hline\\[-2mm] 
\ce{NdCo2}                  & \ce{Nd6O} & 0(0) \\
\ce{NdCo_{1.75}Ni_{0.25}}   & \ce{Nd6O} & 0.3(1) \\
                            & \ce{NdNi} & 3(1) \\
\ce{NdCo_{1.50}Ni_{0.50}}   & \ce{Nd6O} & 0.2(1) \\
                            & \ce{NdNi} & 1.5(4) \\
\ce{NdCo_{1.25}Ni_{0.75}}   & \ce{Nd6O} & 0.3(1) \\
                            & \ce{NdNi} & 1.5(6) \\
\ce{NdCoNi}                 & \ce{Nd(CoNi)3} & 2.1(5) \\[1mm]
\end{tabular}
\end{ruledtabular}
\label{tab:EDS}
\end{table}

\section{SR-PXD Data}
\label{sec:XRD}

\begin{figure*}[htbp]
    \centering
    \includegraphics[width = 0.5\textwidth]{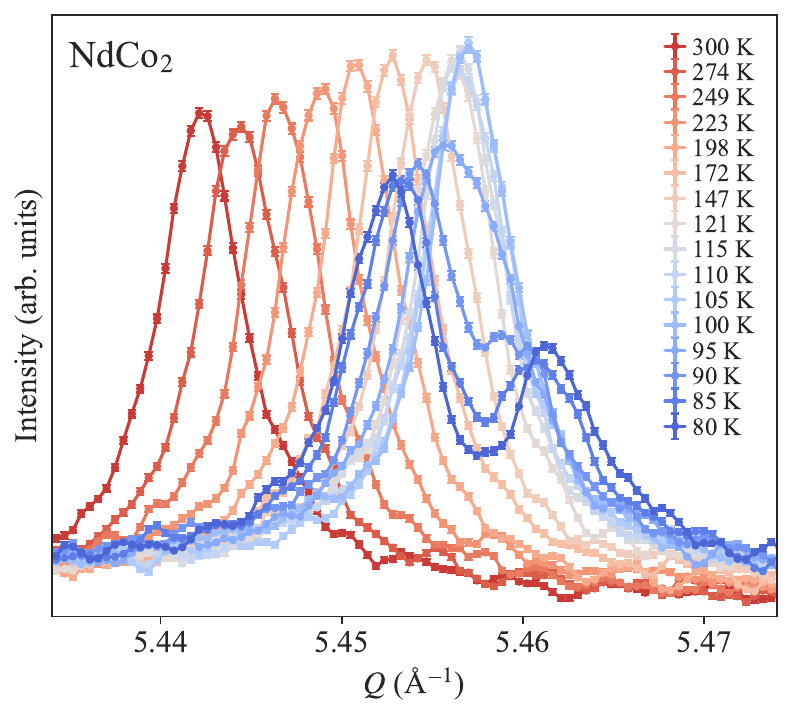}
    \caption{SR-PXD data obtained of the (620) Bragg peak of \ce{NdCo2} at temperatures ranging from 80 to 300~K. The sample was cooled down using a nitrogen blower and measured during heating. Below 100~K, peak asymmetry is observed, and below 95~K, the peak is clearly split. $\lambda$ = 0.71859~{\AA}.}
    \label{fig:NdCo2_LowT_XRD}
\end{figure*}
\newpage
\begin{figure*}[htbp]
    \centering
    \includegraphics[width = \textwidth]{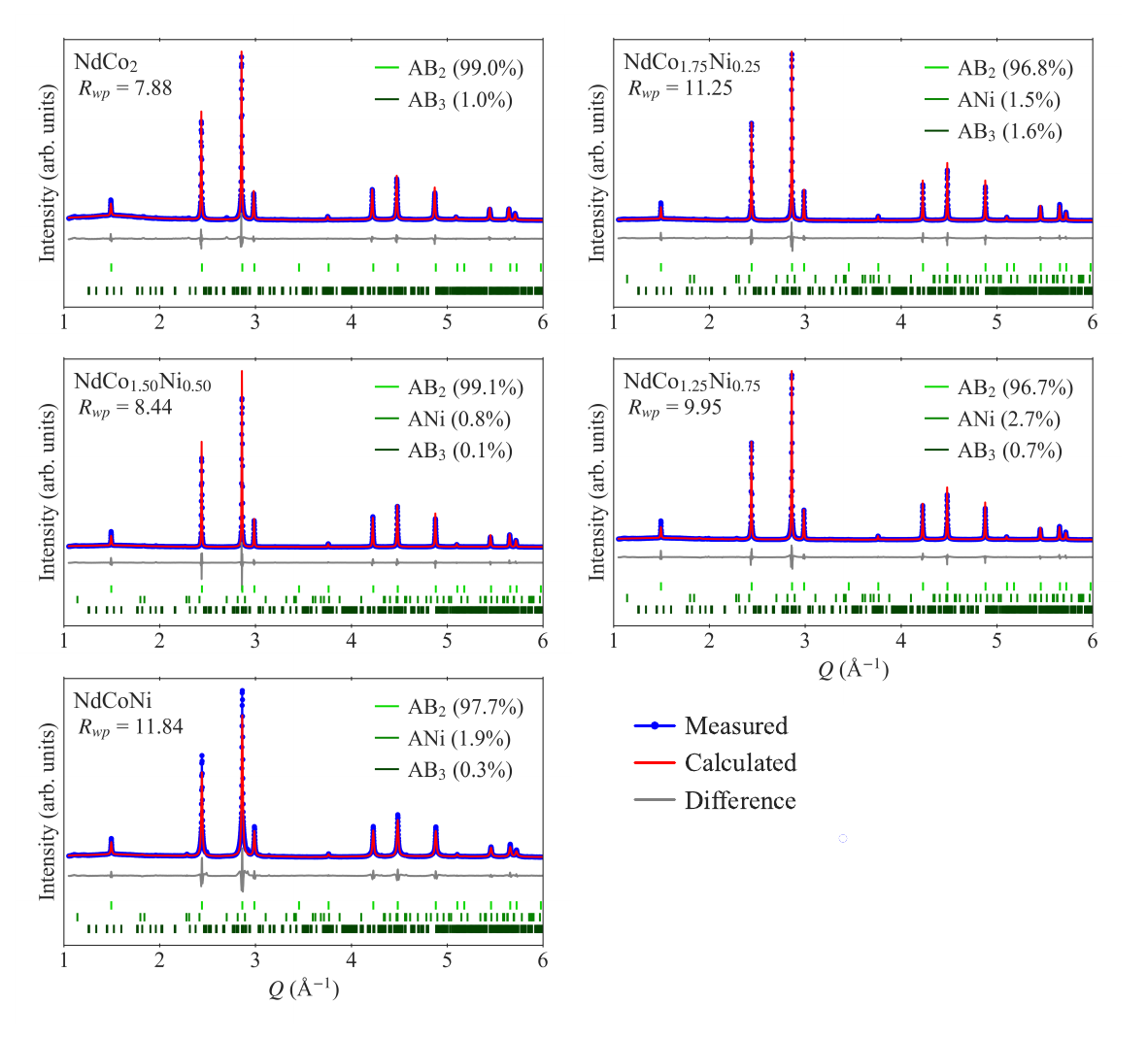}
    \caption{Rietveld refinement (red lines) of SR-PXD data (blue circles) obtained at BM01 of all compounds at room temperature using $\lambda$ = 0.49537~{\AA}. Grey lines indicate the difference between the measured data and the refinement. Vertical bars indicate Bragg reflections for the main \ce{AB2} cubic Laves phase (C15), as well as the secondary \ce{AB} (space group $Cmcm$) and \ce{AB3} (space group $Pnma$) phases with their respective weight percentages as determined from the refinement. A is Nd, and B is the stoichiometric transition metal composition for the given compound.}
    \label{fig:Refined_XRD}
\end{figure*}

\newpage
\section{Neutron Scattering}
\label{sec:PND}

\begin{figure*}[htbp]
  \centering
  {\includegraphics[width = 0.49\textwidth]{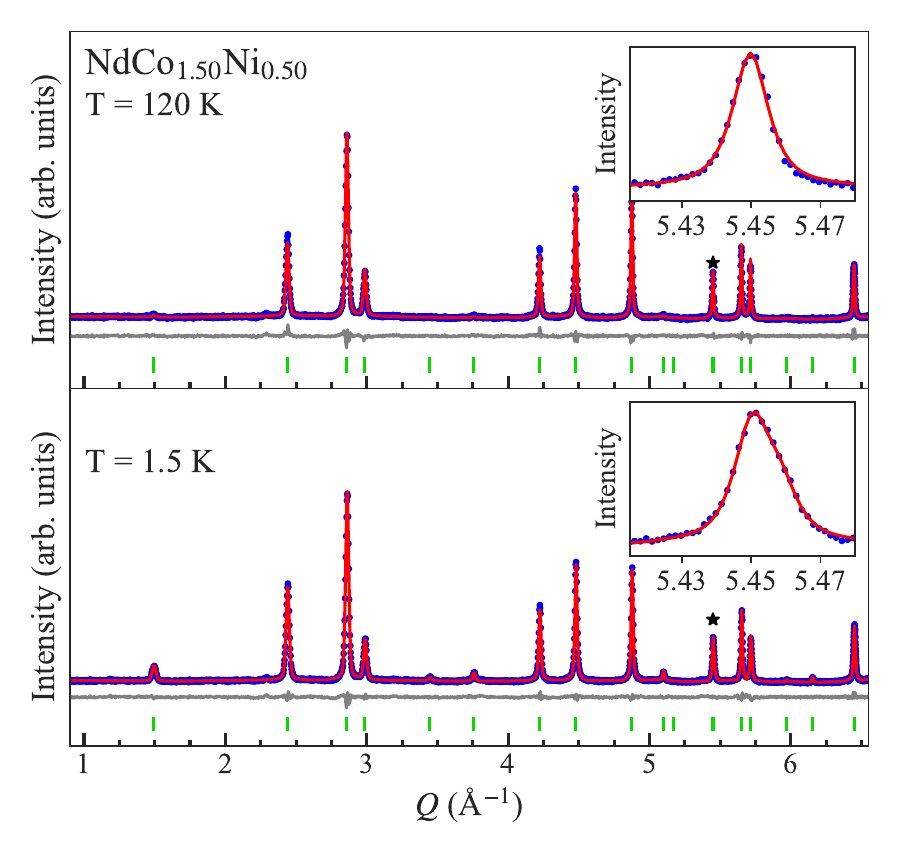} \label{fig:Refined_PSI_NdCo1.50Ni0.50}}
  {\includegraphics[width = 0.49\textwidth]{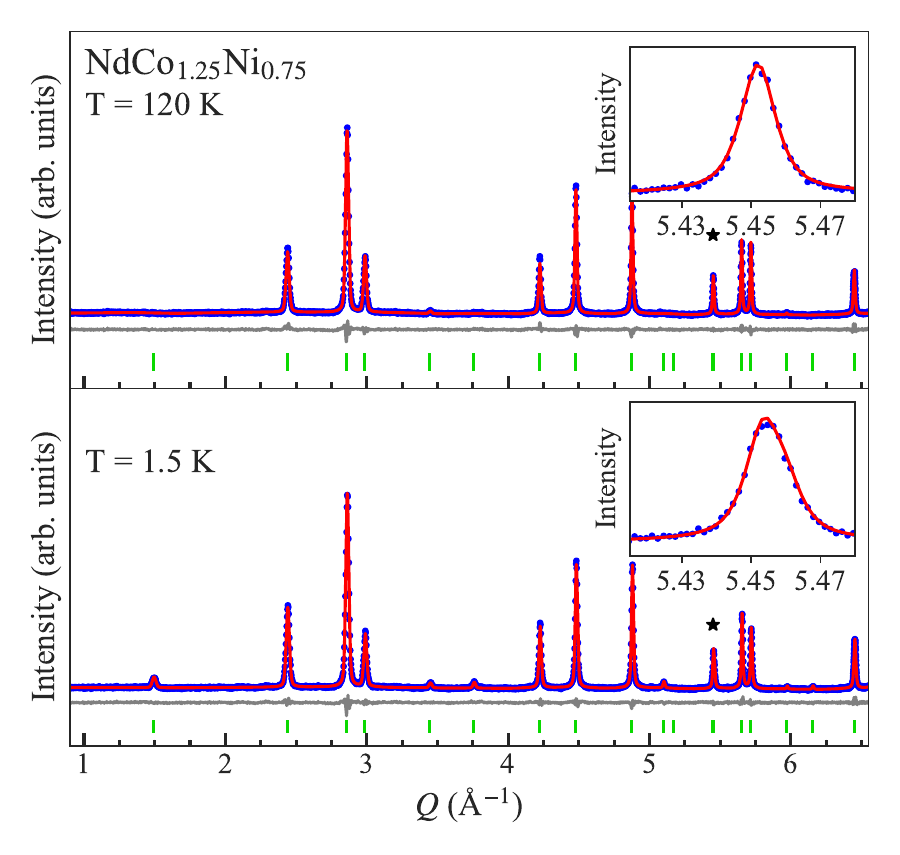} \label{fig:Refined_PSI_NdCo1.25Ni0.75}} \\
  {\includegraphics[width = 0.49\textwidth]{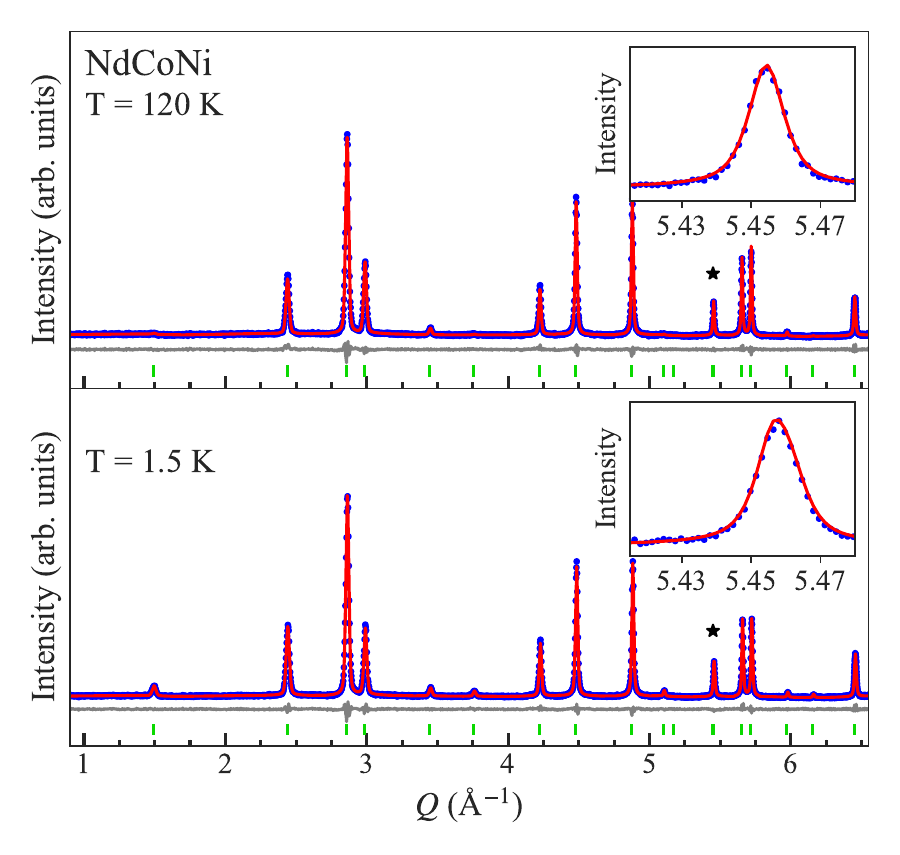} \label{fig:Refined_PSI_NdCoNi}} \\
  \caption{Experimental (blue circles) and calculated diffraction patterns (red lines), and difference (grey lines) for Rietveld refinements of PND data of \ce{NdCo_{1.50}Ni_{0.50}}, \ce{NdCo_{1.25}Ni_{0.75}}, and \ce{NdCoNi} in a cubic (120~K, space group \ce{\textit{Fd}\Bar{3}\textit{m}}) and tetragonal (45~K, space group $I4_1/amd$) phase, using $\lambda = 1.886$~{\AA}. Error bars are within the size of the markers. Vertical bars indicate Bragg reflections for the refined \ce{AB2} phase. The inset shows the (620) Bragg peak, marked with a star in the plot.}
  \label{fig:Refined_PSI}
\end{figure*}

\newpage

\begin{table}[htbp]
\centering 
\caption{Results from the refinements of the PND data using Mag2Pol. The first part of the table shows the results for the compounds in their cubic state at 120~K. The second part shows the results for the compounds in their tetragonal states: 45~K for \ce{NdCo2} and \ce{NdCo_{1.75}Ni_{0.25}}, and 1.5~K for the remaining compounds. The last part of the table shows the results for \ce{NdCo2} and \ce{NdCo_{1.75}Ni_{0.25}} in their orthorhombic state at 1.5~K. The magnetic moments are refined along the $c$ axis in the tetragonal phase and along the $b$ axis in the orthorhombic phase.}
\begin{ruledtabular}
\begin{tabular}{l c c c c c}
120~K: Cubic & \ce{NdCo2} & \ce{NdCo_{1.75}Ni_{0.25}} & \ce{NdCo_{1.50}Ni_{0.50}} & \ce{NdCo_{1.25}Ni_{0.75}} & \ce{NdCoNi} \\[1mm] 
\hline \\[-2mm]
$a$ ({\AA}) & 7.2832(2) & 7.27837(3) & 7.2764(3) & 7.2745(3) & 7.27066(2) \\
$V$ ({\AA}$^3$) & 386.34(4) & 385.56(5) & 385.26(6) & 384.95(5) & 384.34(3) \\
$B_\text{Nd}$ ({\AA}$^2$) & 0.28(2) &  0.53(2) & 0.44(2) & 0.47(2) & 0.51(1) \\
$B_\text{Co/Ni}$ ({\AA}$^2$) & 0.45(3) & 0.67(3) & 0.78(3) & 0.71(2) & 0.64(2) \\
$\chi^2$ & 2.22 & 2.12 & 1.97 & 2.18 & 2.21 \\[1mm] 
\hline \\[-3mm]
\hline \\[-2mm]
45~K / 1.5~K: Tetragonal & \ce{NdCo2} & \ce{NdCo_{1.75}Ni_{0.25}} & \ce{NdCo_{1.50}Ni_{0.50}} & \ce{NdCo_{1.25}Ni_{0.75}} & \ce{NdCoNi} \\[1mm] 
\hline \\[-2mm]
$a$ ({\AA}) & 5.15719(6) & 5.1467(3) & 5.1450(2) & 5.1428(2) & 5.1394(3) \\
$c$ ({\AA}) & 7.2659(1) & 7.2684(3) & 7.2636(2) & 7.2627(3) & 7.2600(3) \\
$V$ ({\AA}$^3$) & 193.24(1) & 192.53(5) & 192.27(4) & 192.08(4) & 191.76(5) \\
$B_\text{Nd}$ ({\AA}$^2$) & 0.18(2) & 0.36(2) & 0.31(2) & 0.28(1) & 0.32(1) \\
$B_\text{Co/Ni}$ ({\AA}$^2$) & 0.31(3) & 0.41(3) & 0.31(2) & 0.33(2) & 0.35(1) \\
$M_\text{Nd}$ ($\mu_\text{B}$) & 2.53(3) & 2.30(4) & 2.72(3) & 2.61(3) & 2.36(3) \\
$M_\text{Co/Ni}$ ($\mu_\text{B}$) & 0.75(2) & 0.55(3) & 0.63(3) & 0.63(3) & 0.51(2) \\
$\chi^2$ & 1.86 & 1.92 & 1.90 & 2.03 & 2.08 \\[1mm] 
\hline \\[-3mm]
\hline \\[-2mm]
1.5~K: Orthorhombic & \ce{NdCo2} & \ce{NdCo_{1.75}Ni_{0.25}} & - & - & - \\[1mm] 
\hline \\[-2mm]
$a$ ({\AA}) & 5.1504(5) & 5.1476(4) & - & - & - \\
$b$ ({\AA}) & 5.1427(5) & 5.1385(4) & - & - & - \\
$c$ ({\AA}) & 7.2965(5) & 7.2832(4) & - & - & - \\
$V$ ({\AA}$^3$) & 193.26(5) & 192.64(4) & - & - & - \\
$z_\text{Nd}$ & 0.3739(8) & 0.3727(5) & - & - & - \\
$B_\text{Nd}$ ({\AA}$^2$) & 0.13(2) & 0.26(2) & - & - & - \\
$B_\text{Co/Ni}$ ({\AA}$^2$) & 0.23(3) & 0.25(3) & - & - & - \\
$M_\text{Nd}$ ($\mu_\text{B}$) & 2.67(6) & 2.80(4) & - & - & - \\
$M_\text{Co/Ni}$ ($\mu_\text{B}$) & 0.75(4) & 0.63(2) & - & - & - \\
$\chi^2$ & 1.08 & 1.16 & - & - & - \\[1mm] 
\end{tabular}
\end{ruledtabular}
\label{tab:PND_refinement}
\end{table}

\newpage

\begin{figure*}[htbp]
    \centering
    \includegraphics[width = 0.85\textwidth]{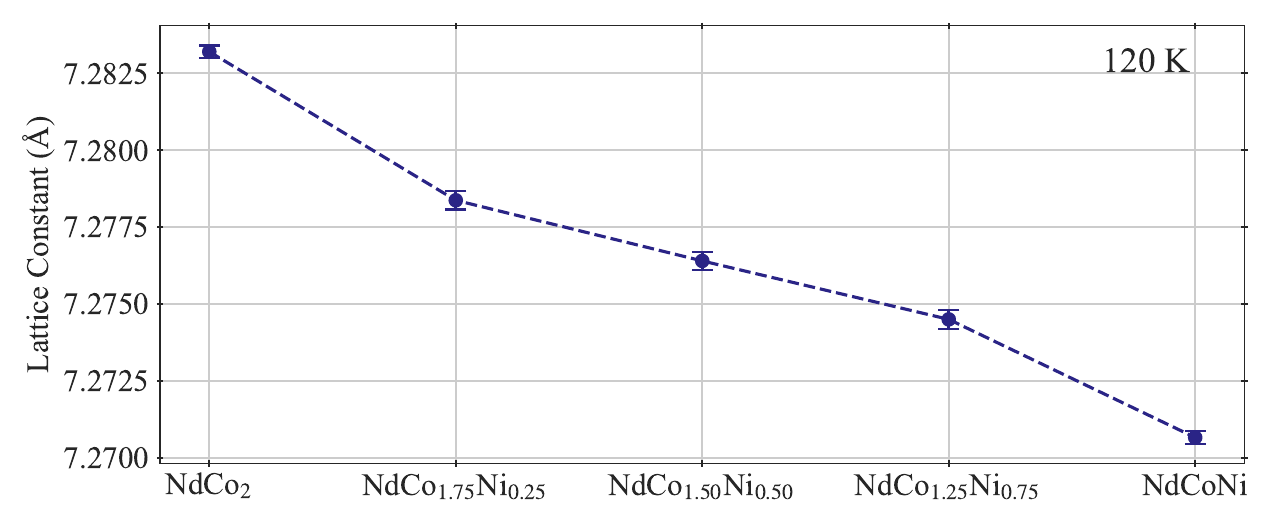}
    \caption{Lattice constants in {\AA} of the cubic unit cell axis $a$ obtained by Rietveld refinements of the C15 cubic Laves phase using PND data at 120~K using Mag2Pol. Error bars indicate the error of the refinement. Values are tabulated in Table~\ref{tab:PND_refinement}.}
    \label{fig:Lattice_Constants}
\end{figure*}

\begin{figure*}[htbp]
    \centering
    {\includegraphics[width = \textwidth]{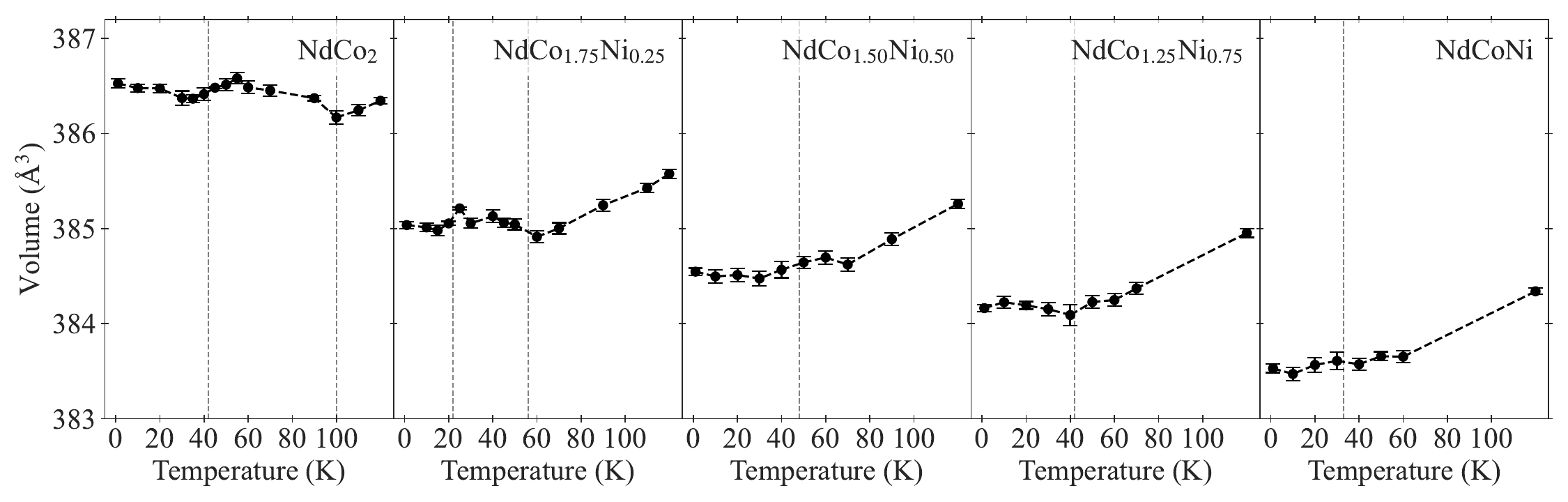}\label{fig:Volume_PND}} \\
    \caption{Unit cell volume ({\AA}$^3$) obtained by Rietveld refinements of PND data at selected temperatures in the temperature interval from 1.5 to 120~K. The error bars indicate the error of the refinements. Vertical gray lines indicate the phase transitions. The pseudo-cubic volume is shown for the tetragonal and orthorhombic structures, which are twice those of the corresponding tetragonal and orthorhombic unit cell volumes, respectively.}
    \label{fig:Refined_lattice_volume}
\end{figure*}

\newpage

\begin{figure*}[h!]
    \centering
    \includegraphics[width = 0.55\textwidth]{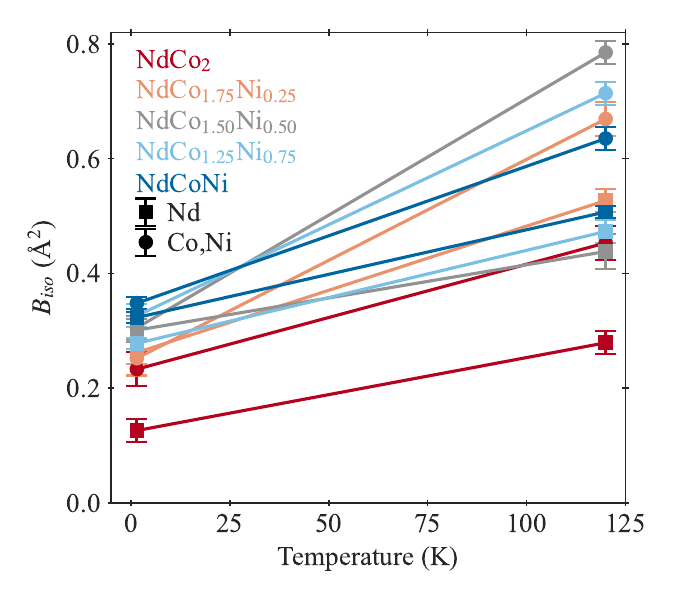}
    \caption{Thermal displacement parameter $B_{iso}$, determined at 1.5 and 120~K through refinements of PND data, while interpolated values using these two data points were used for intermediate temperatures. The error bars indicate the calculated error of the refinements. Tabulated values are given in Table~\ref{tab:PND_refinement}.}
    \label{fig:B}
 \end{figure*}

  \begin{figure*}[h!]
    \centering
    \includegraphics[height = 3.7cm]{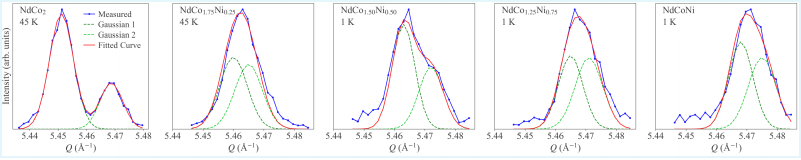}
    \includegraphics[height = 3.7cm]{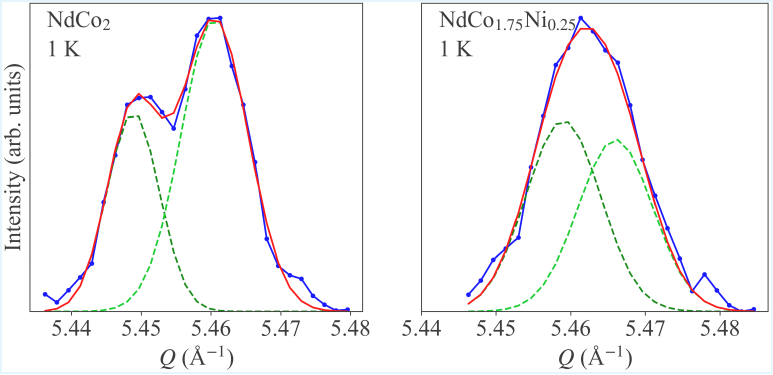}
    \caption{Curve fitting using Gaussian functions of the (620) Bragg peak at 1 and 45~K measured using PND. The first row shows the data of the tetragonal structures, and the second row shows the data of the orthorhombic structures.}
    \label{fig:Fitted_PND}
 \end{figure*}

\begin{figure*}[htbp]
  \centering
  \subfloat[][(111) Bragg peak]{\includegraphics[width = 0.99\textwidth]{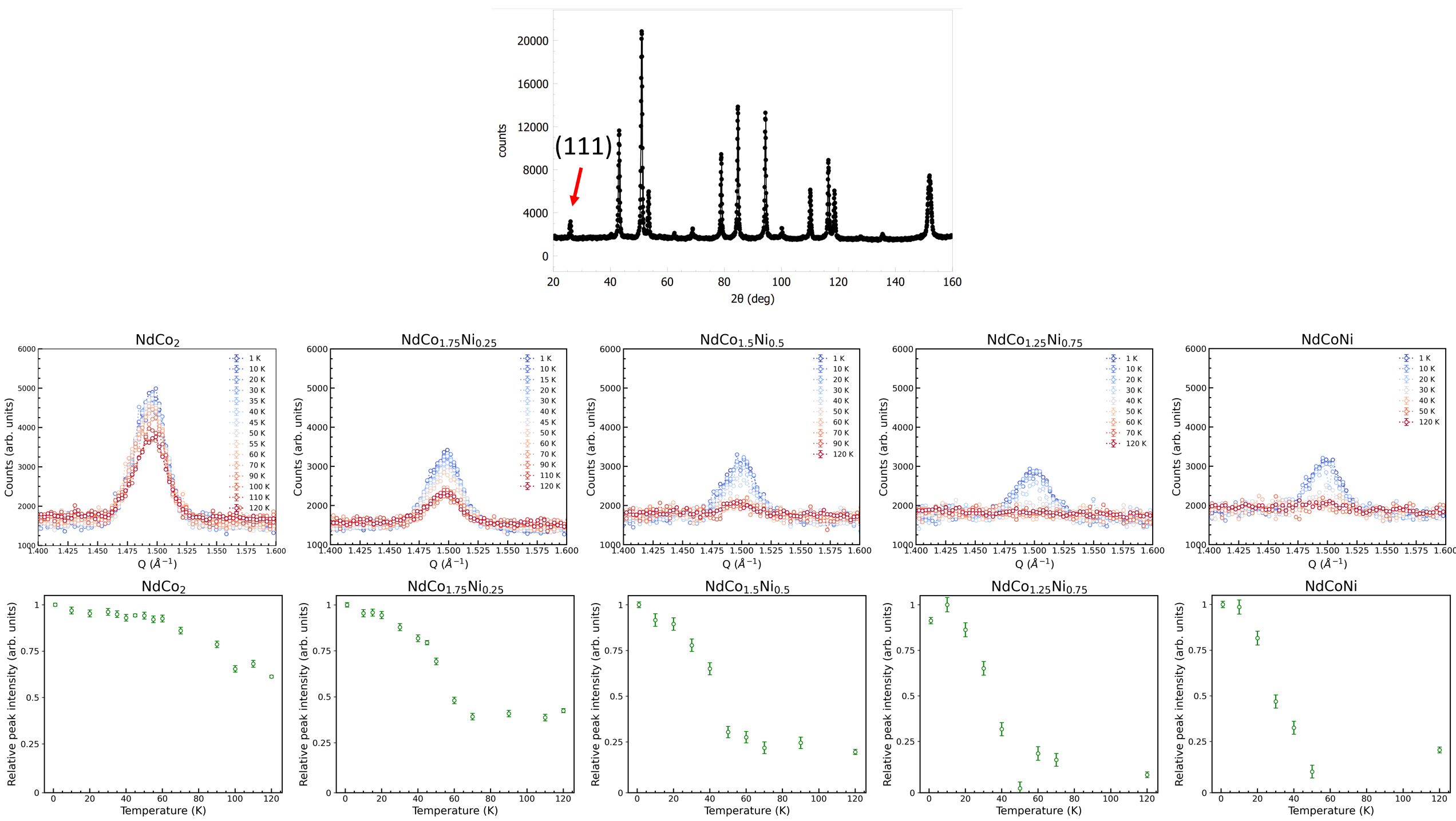} \label{fig:Peak_111}} \\
  \subfloat[][(620) Bragg peak]{\includegraphics[width = 0.99\textwidth]{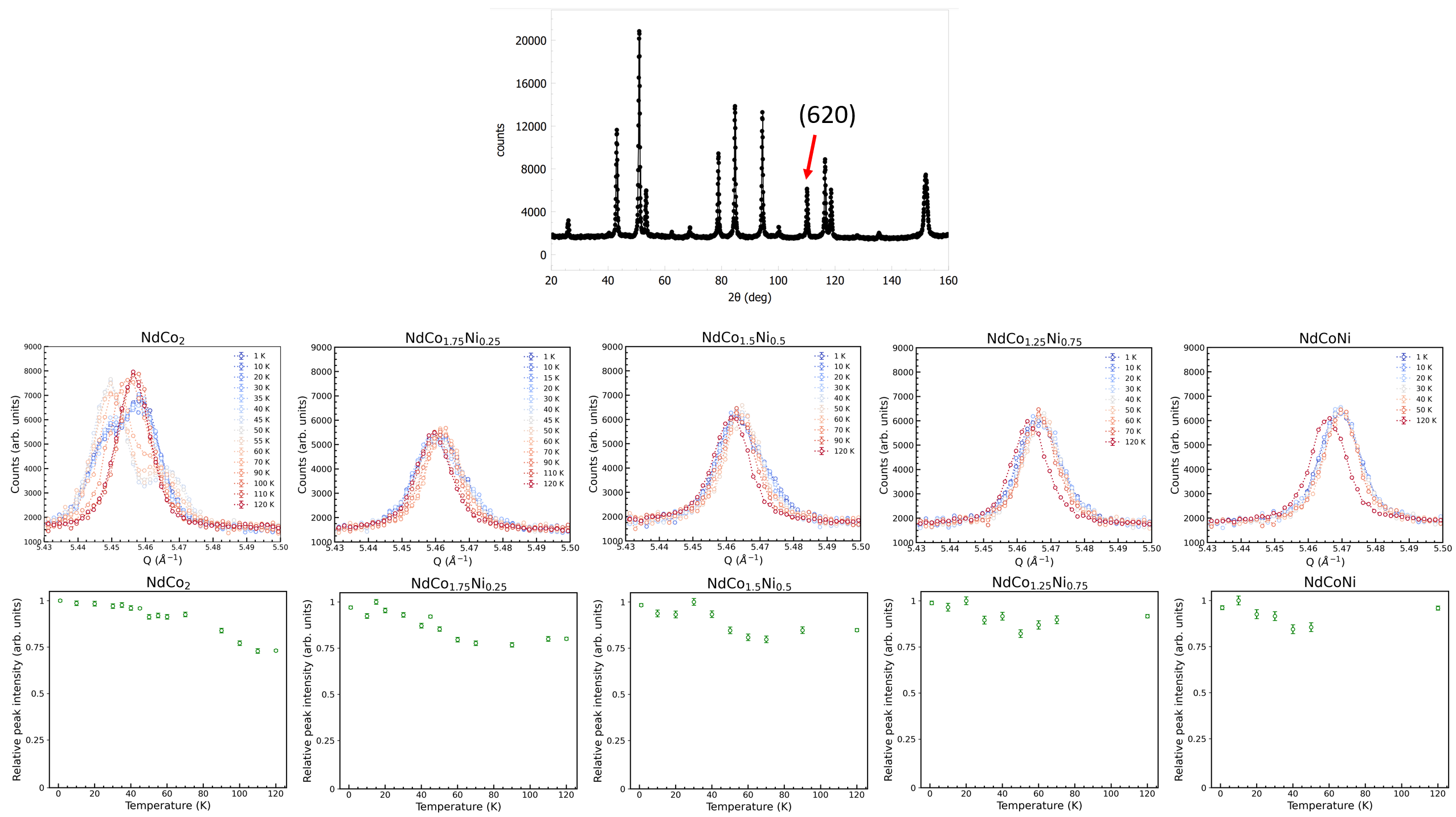} \label{fig:Peak_620}} \\
  \caption{Intensities and integrated peak intensities of the (a) (111) and (b) (620) Bragg peaks for all the five \ce{NdCo_{2-x}Ni_x} (0 $\leq$ x $\leq$ 1) compounds at selected temperatures in the range from 1.5 to 120~K. The integrated intensities are approximated by subtracting the background around the given peaks. The low-$Q$ (111) peak shows a strong correlation with the magnetic moment, and a drop in the integrated peak intensity can be seen upon heating above $T_\text{C}$. The high-$Q$ (620) peak was used to observe the tetragonal and orthorhombic distortions at low temperatures.}
  \label{fig:Peaks_neutron}
\end{figure*}

\newpage
\section{Bulk Magnetization, Magnetocaloric, and Heat Capacity Measurements}
\label{sec:PPMS}

\subsection{Detailed Indirect Magnetocaloric Measurement Methodology}

\begin{figure*}[htbp]
  \centering
  {\includegraphics[width = 0.49\textwidth]{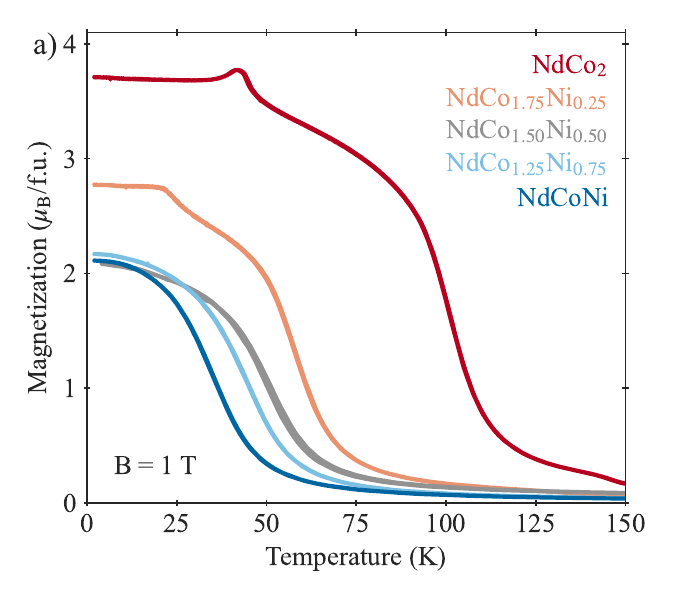} \label{fig:MT_1T}}
  {\includegraphics[width = 0.49\textwidth]{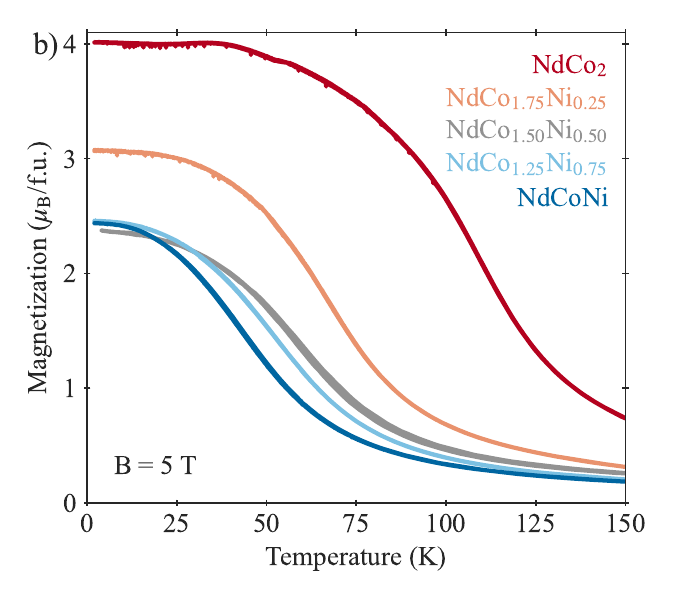} \label{fig:MT_5T}} \\
  \vspace{-0.3cm} 
  {\includegraphics[width = 0.49\textwidth]{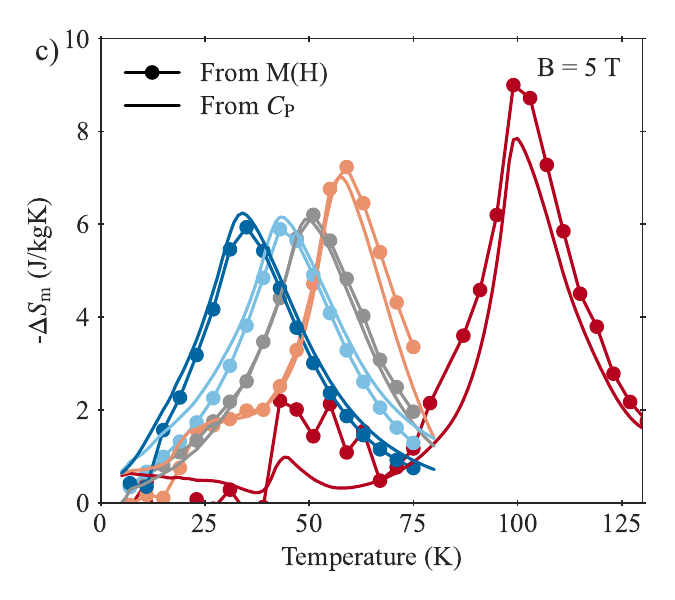} \label{fig:dS}}
  {\includegraphics[width = 0.49\textwidth]{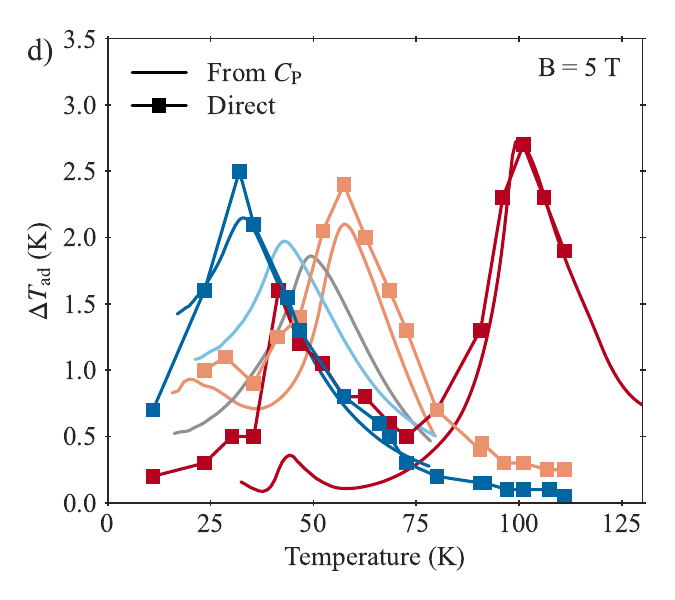} \label{fig:dT}} \\
  \vspace{-0.3cm} 
  {\includegraphics[width = 0.49\textwidth]{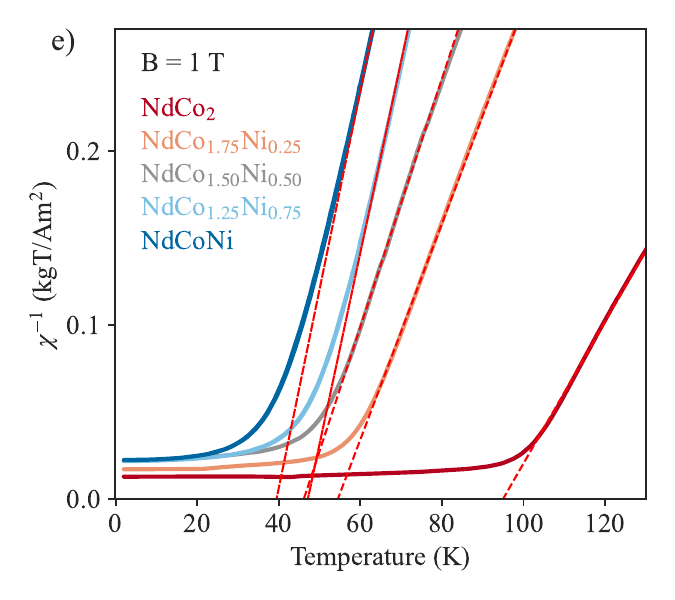} \label{fig:Curie-Weiss}} \\
  \vspace{-0.3cm} 
  \caption{Temperature-dependent bulk magnetization measured at (a) 1~T and (b) 5~T, respectively. (c) $\Delta S_\text{m}$ and (d) $\Delta T_\text{ad}$ obtained at 5~T from $M(H)$ curves, S-T diagrams from $C_\text{P}$ data, and direct pulsed-field measurements. (e) Curie-Weiss fitting if the inverse susceptibility $\chi^{-1}$ at 1~T, used to determine $\mu_\text{cal}$ and $\theta_\text{CW}$.}
  \label{fig:Magnetic}
\end{figure*}

This section presents a detailed description of the methodology for the measurement and analysis of the data used to determine the magnetocaloric effect indirectly in this work. $\Delta S_\text{m}$ and $\Delta T_\text{ad}$ were determined using three different methods ($C_\text{P}$, $M(H)$, and directly), as presented together in Fig.~\ref{fig:Magnetic} c) and d).

\begin{figure}[htbp]
  \centering
  \subfloat[][]{\includegraphics[height = 11 cm]{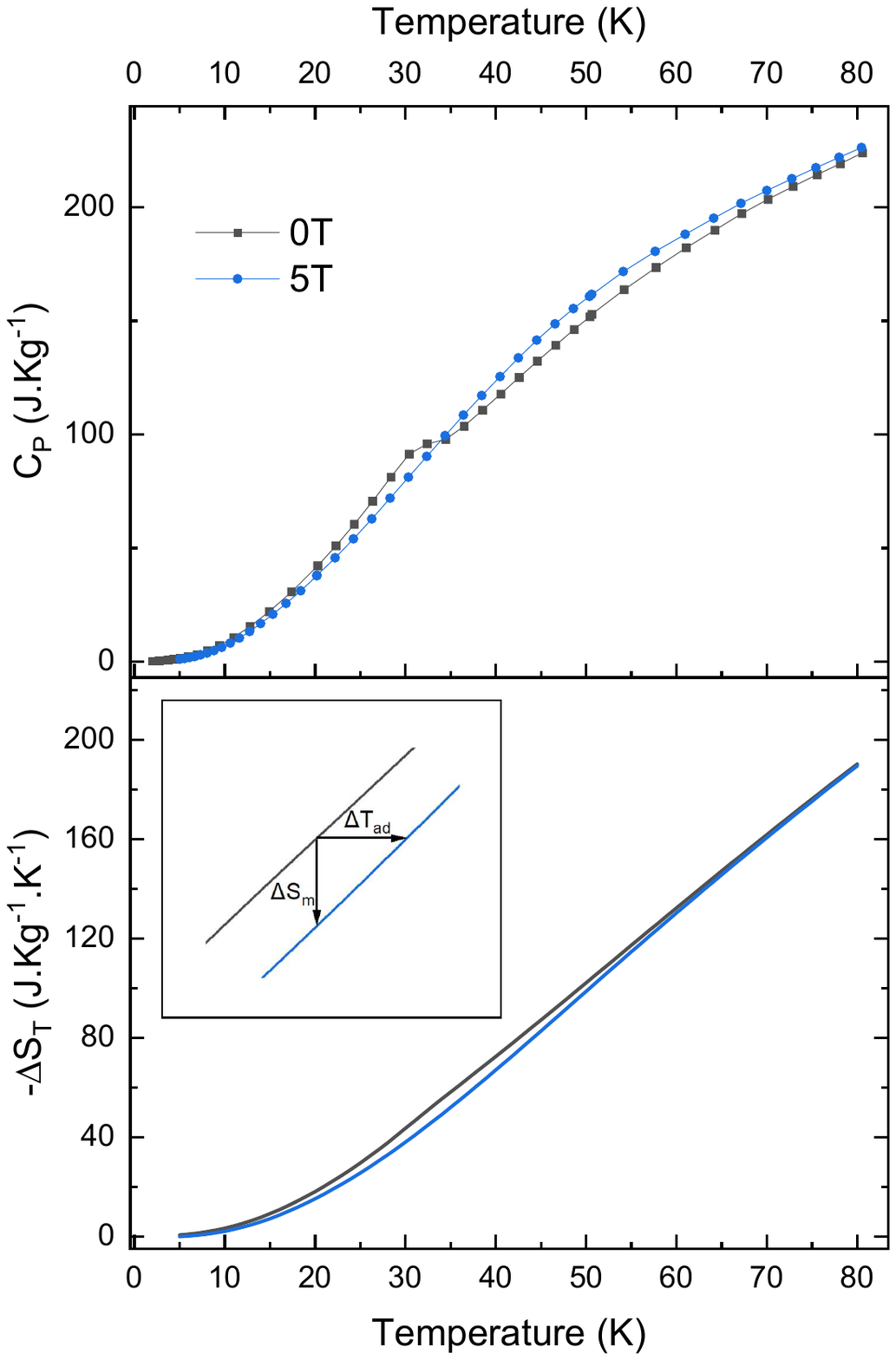} \label{fig:dS_dT_Cp_NdCoNi}}
  \subfloat[][]{\includegraphics[height = 11 cm]{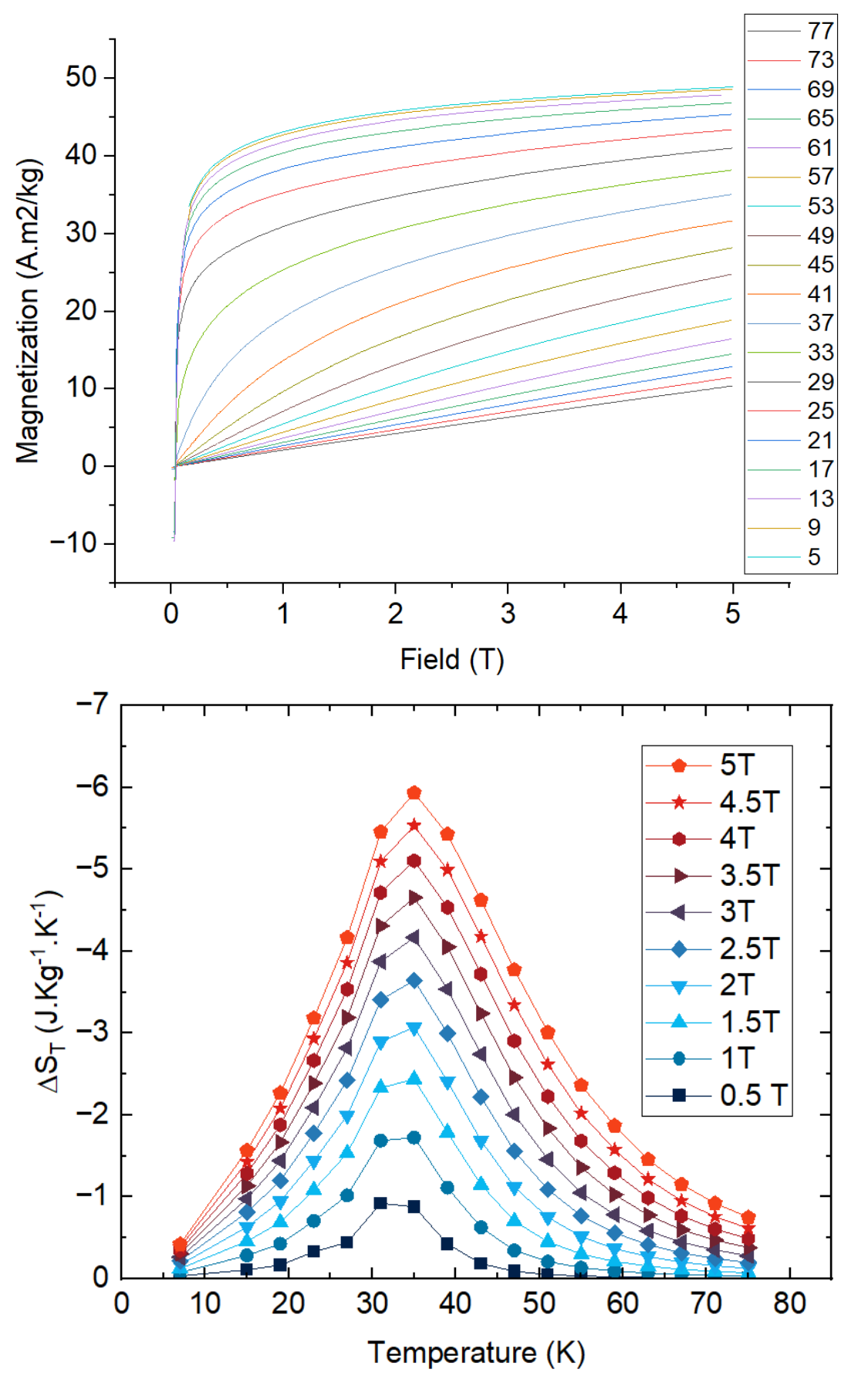} \label{fig:MH_dS}} \\
  \caption{Illustration of the determination of $\Delta S_\text{m}$ and $\Delta T_\text{ad,ind}$ on NdCoNi data. (a) $C_\text{P}$ data (upper) used to find the total entropy (lower) at 0 and 5~T. The vertical and horizontal distances between these determine $\Delta S_\text{m}$ and $\Delta T_\text{ad}$, respectively. (b) $M(H)$ curves (upper) used to determine $\Delta S_\text{m}$ (lower) at different fields.} \label{fig:dS_determination}
\end{figure}

\textbf{From \bm{$C_\text{P}$}}: Illustrated in Fig.~\ref{fig:dS_determination}(a). $C_\text{P}$ measurements provide iso-magnetic entropy $S$ versus temperature $T$ diagrams. This is obtained by integrating $C_\text{P}/T$ over temperature for each field, in this work: $B$=0~T and $B$=5~T. From these curves, we can calculate $\Delta S_\text{m}$ and $\Delta T_\text{ad}$ because:

\begin{align}
    \Delta S_\text{m}(T,\Delta H) &= S(T, H_F) - S(T, H_I)
	\label{eq:entropy},
    \\
    \Delta T_\text{ad}(T,\Delta H) &= T(S, H_F) - T(S, H_I).
	\label{eq:temperature}
\end{align}

\textbf{From \bm{$M(H)$}}:  Illustrated in Fig.~\ref{fig:dS_determination}(b). Field-dependent magnetization curves $M(H)$ were measured at constant temperatures from 0 to 5~T in 4~K increments. From these curves, $\Delta S_\text{m}$ is calculated:

\begin{equation}
    	\Delta S_\text{m}(T,\Delta H) = \mu_0 \int_{H_I}^{H_F} \left(\frac{\partial M(T,H)}{\partial T}\right)_{p,H} dH.
	\label{eq:entropy_2}
\end{equation}

\newpage

\begin{figure}[htbp]
  \centering
  {\includegraphics[width = 0.49\textwidth]{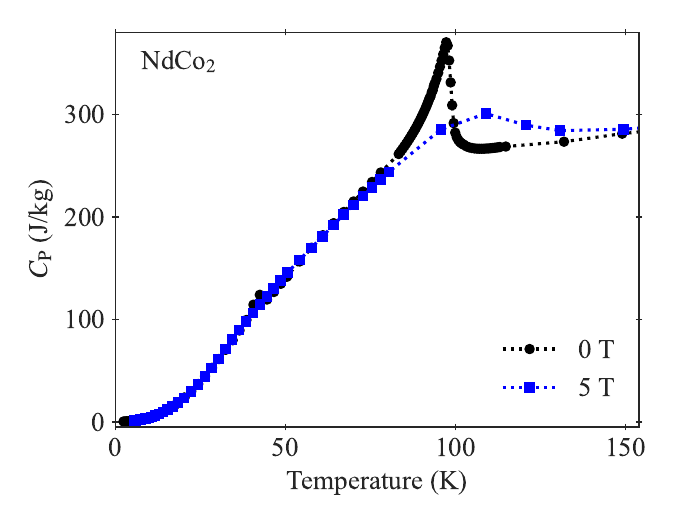} \label{fig:NdCo2_Cp}} \\ \vspace{-0.3cm} 
  {\includegraphics[width = 0.49\textwidth]{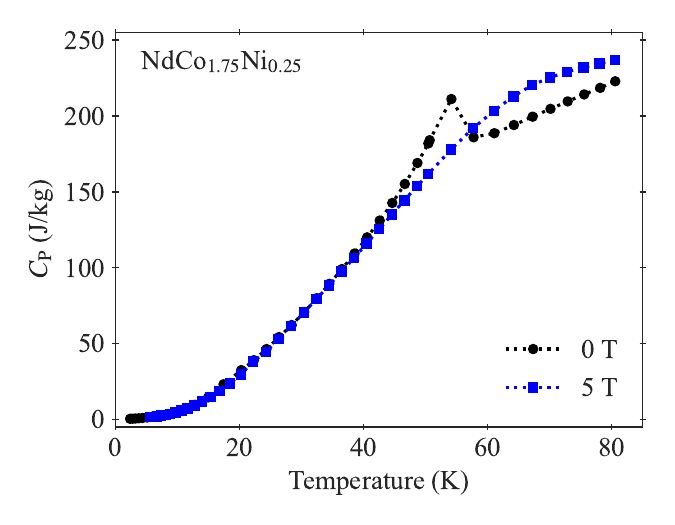} \label{fig:NdCo1.75Ni0.25_Cp}} 
  {\includegraphics[width = 0.49\textwidth]{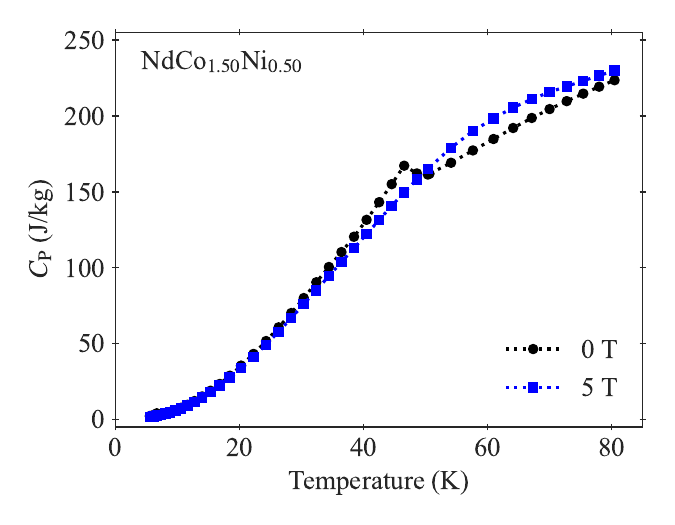} \label{fig:NdCo1.50Ni0.50_Cp}} \\ \vspace{-0.3cm} 
  {\includegraphics[width = 0.49\textwidth]{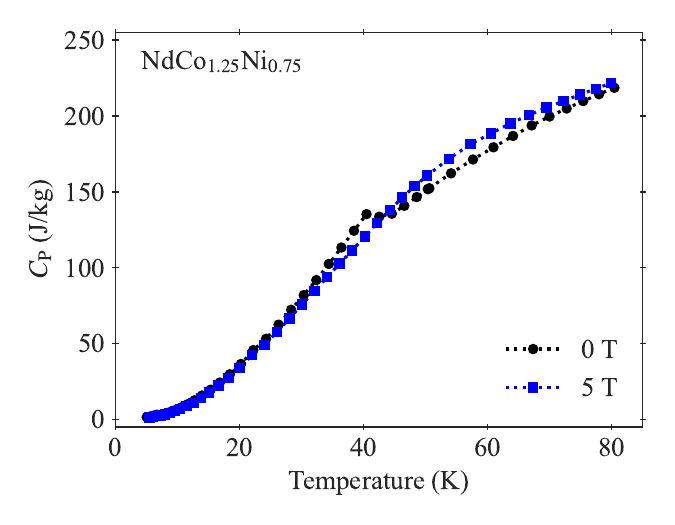} \label{fig:NdCo1.25Ni0.85_Cp}}
  {\includegraphics[width = 0.49\textwidth]{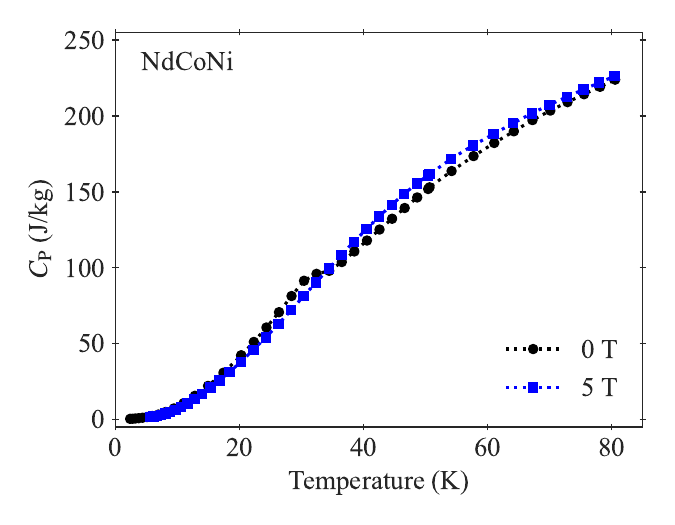} \label{fig:NdCoNi_Cp}} \\
  \caption{The markers show the heat capacity $C_\text{P}$ data measured at 0 and 5~T with the $2\tau$ approach. Lines connect data points to guide the eye.} \label{fig:Cp}
\end{figure}
\newpage
\newpage
\begin{figure}[htbp]
  \centering
  {\includegraphics[width = 0.49\textwidth]{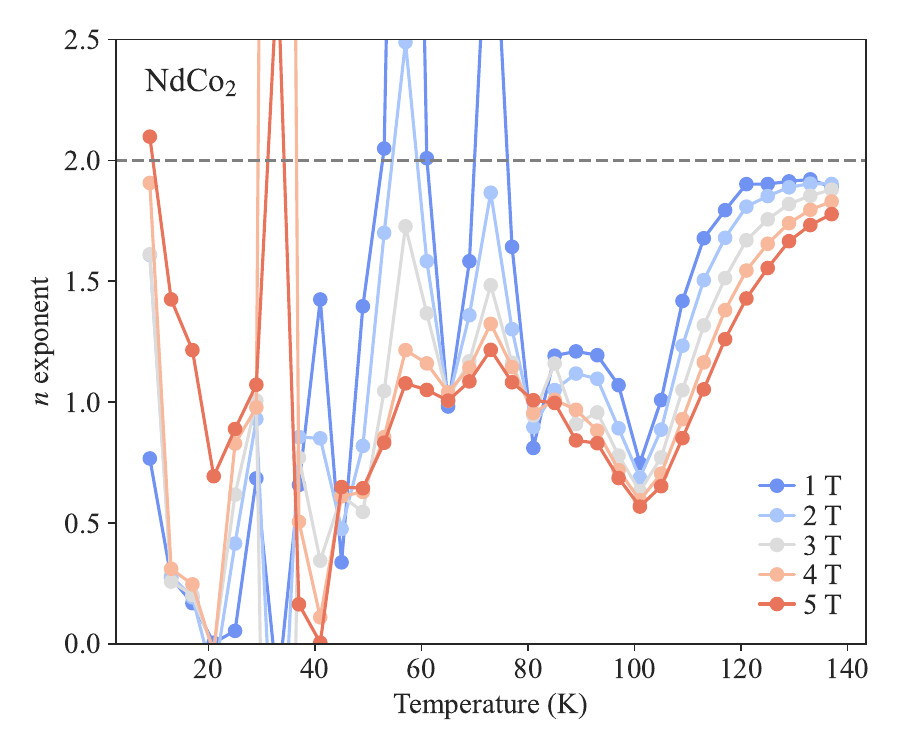} \label{fig:NdCo2_n_exponent}} \\ \vspace{-0.3cm} 
  {\includegraphics[width = 0.49\textwidth]{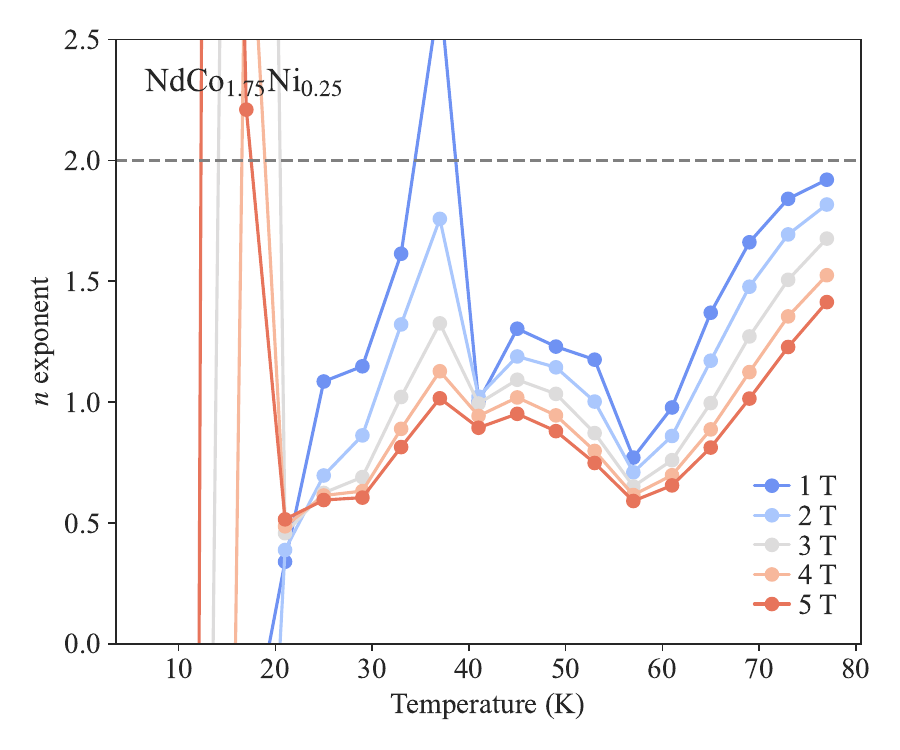} \label{fig:NdCo1.75Ni0.25_n_exponent}} 
  {\includegraphics[width = 0.49\textwidth]{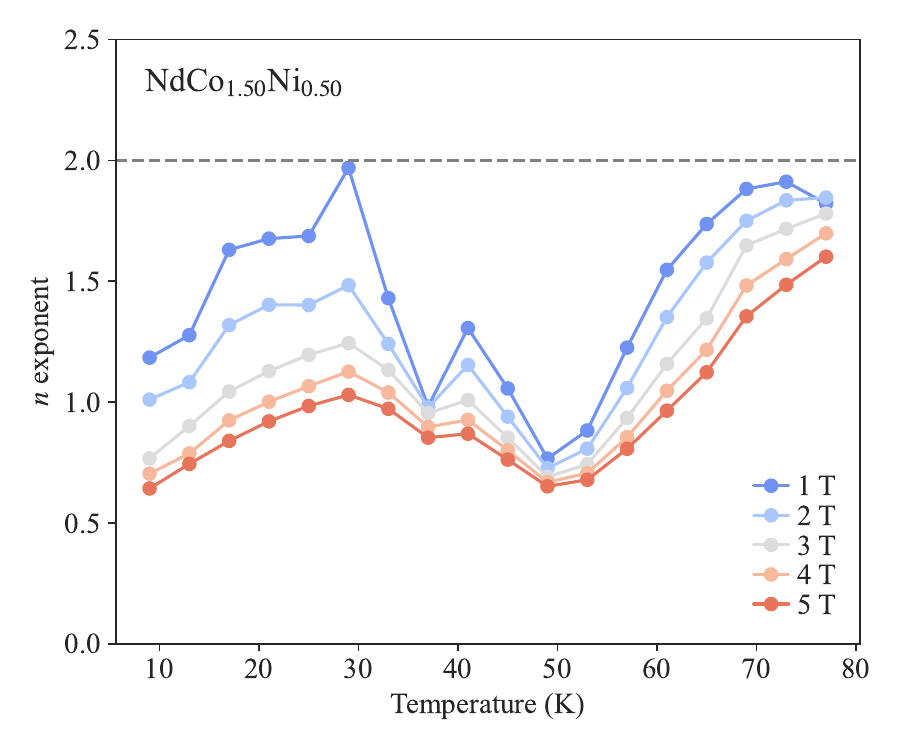} \label{fig:NdCo1.50Ni0.50_n_exponent}} \\ \vspace{-0.3cm} 
  {\includegraphics[width = 0.49\textwidth]{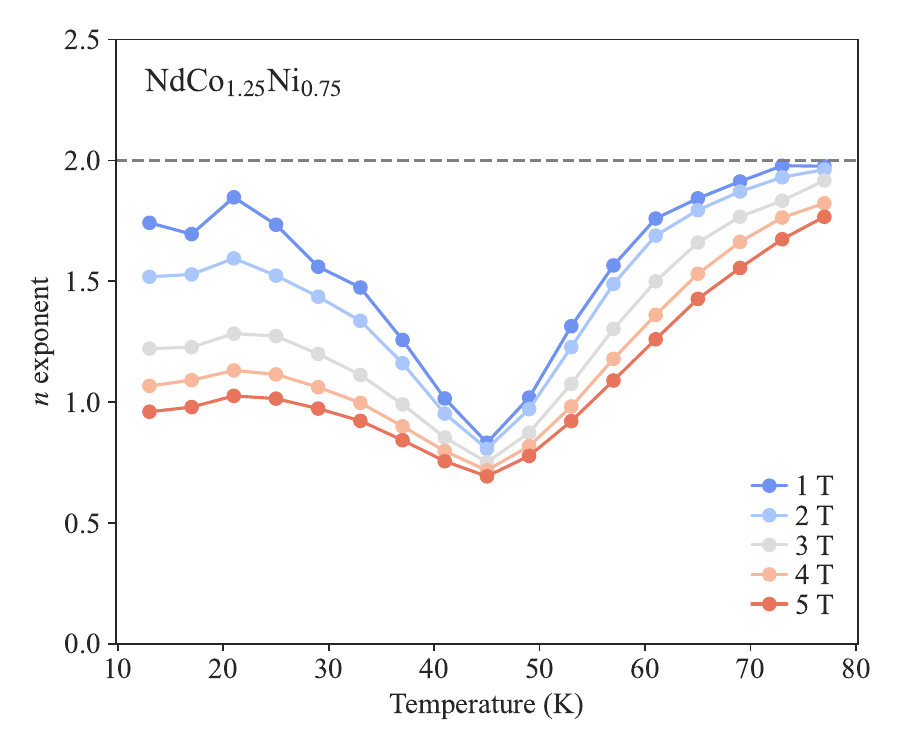} \label{fig:NdCo1.25Ni0.75_n_exponent}}
  {\includegraphics[width = 0.49\textwidth]{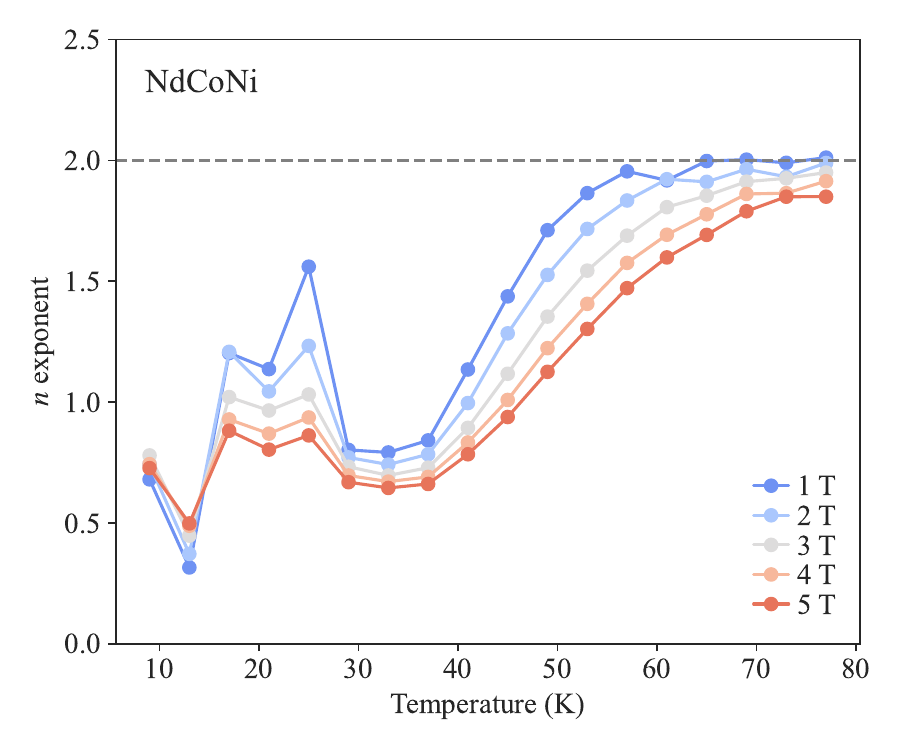} \label{fig:NdCoNi_n_exponent}} \\
  \caption{$n$ exponent analysis of $M(H)$ data shown for different magnetic fields between 1 and 5~T. $n = 2$ is marked with a horizontal dotted line since an $n$ exceeding 2 indicates a first-order phase transition. Markers show the data points, while the lines connect these data points to guide the eye.} \label{fig:n_exponent}
\end{figure}

\newpage

\begin{figure*}[htbp]
  \centering
  {\includegraphics[width = \textwidth]{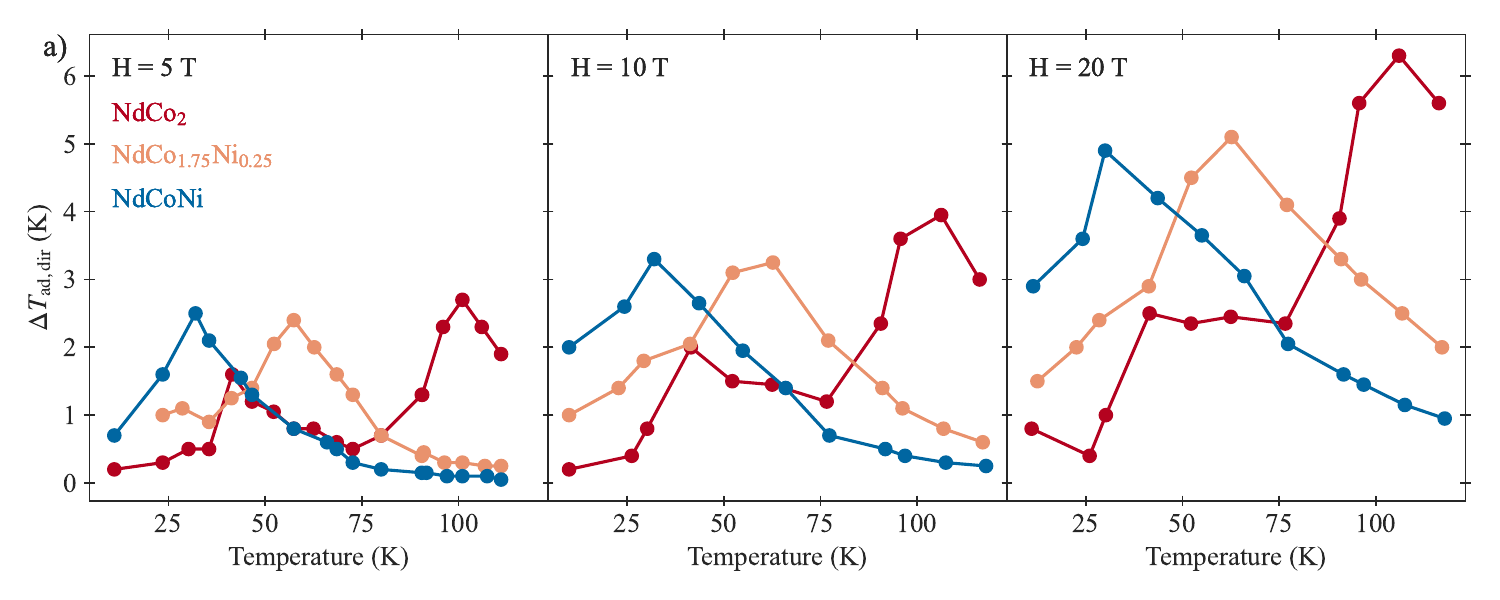} \label{fig:Pulsed_Fields}} \\ 
  {\includegraphics[width = 0.49\textwidth]{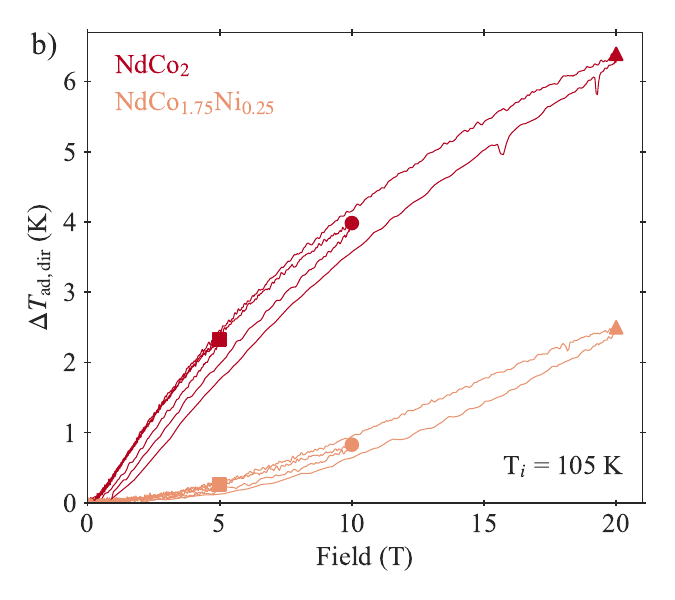} \label{fig:Pulse}} 
  {\includegraphics[width = 0.49\textwidth]{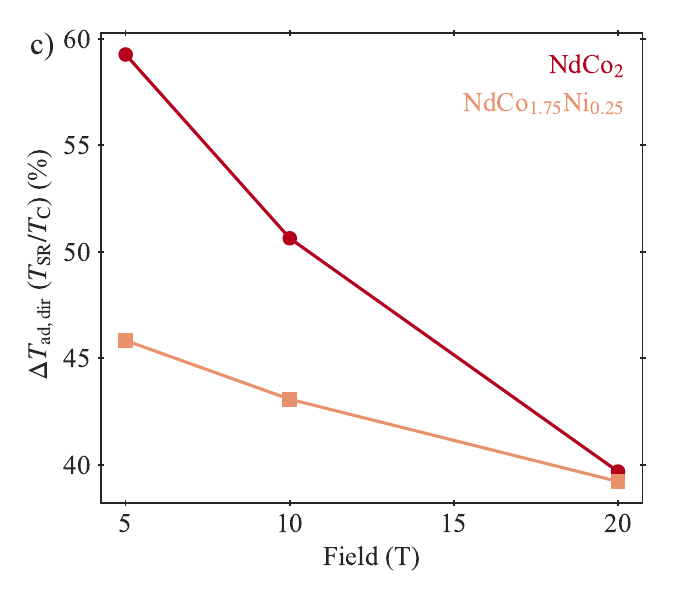} \label{fig:Pulsed_SOPT_FOPT}} \\ 
  \caption{Pulsed-field results obtained at HLD. (a) $\Delta T_\text{ad,dir}$ results using 5, 10, and 20~T. (b) $\Delta T_\text{ad,dir}$ measured at 105~K as a function of magnetic field for pulsed-fields of 5, 10, and 20~T, showing the evolution of these. (c) $\Delta T_\text{ad,dir}$ of the $T_\text{SR}$ (42~K for \ce{NdCo2}, 22~K for \ce{NdCo_{1.75}Ni_{025}}) relative to the $T_\text{C}$ (100~K for \ce{NdCo2}, 56~K for \ce{NdCo_{1.75}Ni_{025}}) as a function of field.}
  \label{fig:dT_direct}
\end{figure*}

\newpage

\begin{table*}[htbp]
\centering 
\caption{Calculated and experimental magnetic properties of the studied compounds. The de Gennes factor $G$ and the theoretical effective magnetic moment $\mu_\text{cal}$ are calculated for Nd. The experimental effective magnetic moment $\mu_\text{eff}$ and the paramagnetic Curie temperature $\theta_\text{CW}$ were determined through Curie–Weiss fitting using a field of 1~T. The saturation magnetization $M_\text{S}$ and the Curie temperature $T_\text{C}$ were determined from temperature- and field-dependent measurements, respectively, using a field of 5~T and 0.01~T, respectively. Remnant magnetization $M_\text{R}$ and coercivity $H_\text{C}$ were determined from field-dependent measurements at 5~K.}
\begin{ruledtabular}
\begin{tabular}{l c c c c c c c c}
Compounds & $G$ & $\mu_\text{cal}$ ($\mu_\text{B}$/f.u.) & $\mu_\text{eff}$ ($\mu_\text{B}$/f.u.) & $M_\text{S}$ ($\mu_\text{B}$/f.u.) & $M_\text{R}$ ($\mu_\text{B}$/f.u.) & $H_\text{C}$ (mT) & $\theta_\text{CW}$ (K) & $T_\text{C}$ (K) \\[2mm] 
\hline \\[-2mm]
\ce{NdCo2}                & 1.84 & 3.62 & 7.17 & 3.99 & 0.19 & 10 & 95 & 100 \\
\ce{NdCo_{1.75}Ni_{0.25}} & 1.84 & 3.62 & 5.81 & 3.05 & 0.47 & 44 & 55 & 56 \\
\ce{NdCo_{1.50}Ni_{0.50}} & 1.84 & 3.62 & 5.42 & 2.37 & 0.56 & 55 & 46 & 48 \\
\ce{NdCo_{1.25}Ni_{0.75}} & 1.84 & 3.62 & 4.36 & 2.30 & 1.17 & 62 & 47 & 42 \\
\ce{NdCoNi}               & 1.84 & 3.62 & 4.28 & 2.42 & 0.80 & 39 & 39 & 34 \\
\end{tabular}
\end{ruledtabular}
\label{tab:magneticprop}
\end{table*}

\begin{table*}[htbp]
\centering 
\caption{Magnetocaloric effect results. $T_\text{tr}$ represents the magnetic transition temperatures, both $T_\text{C}$ and $T_\text{SR}$. $\Delta S_\text{m}$ is determined from $M(H)$ isotherms, and the relative cooling power \textit{RCP} is calculated from $\Delta S_\text{m}$. $\Delta T_\text{ad,ind}$ is determined from $\Delta S_\text{m}$ and $C_\text{P}$, while $\Delta T_\text{ad,dir}$ is measured directly using pulsed magnetic fields. Determining the \textit{RCP} of the $T_\text{SR}$ peaks was problematic due to the small and broad nature of the peaks with tails. Thus, this parameter was only determined at $T_\text{C}$.}
\begin{ruledtabular}
\begin{tabular}{l l c c c l l l}
Parameter & $T_\text{tr}$ (K) & $\Delta S_\text{m}$ (J/kgK) & \textit{RCP} (J/kg) & $\Delta T_\text{ad,ind}$ (K) & \multicolumn{3}{c}{$\Delta T_\text{ad,dir}$ (K)} \\[2mm]
Technique &   & $M(H)$ / $C_\text{P}$ / Direct & $M(H)$ & $M(H)$ / $C_\text{P}$ & \multicolumn{3}{c}{Direct} \\[2mm]
\hline \\[-2mm]
Field  & 0.01~T & 5~T & 5~T & 5~T &  5~T & 10~T & 20~T \\[1mm]
\hline \\[-2mm]
\ce{NdCo2}                  & 100 & 9.0 / 7.8 / 7.8 & 222 & 3.1 / 2.7 & 2.7 & 4.0 & 6.3 \\
                            & 42 & 2.2 / 1.0 / 4.3  & -    & 0.8 / 0.4 & 1.6 & 2.0 & 2.5 \\
\ce{NdCo_{1.75}Ni_{0.25}}   & 56 & 7.2 / 7.0 / 8.0 & 188  & 2.1 / 2.1 & 2.4 & 3.0 & 5.1 \\
                            & 22 & 1.6 / 1.8 / 2.4 & -    & 0.9 / 0.9 & 1.1 & 1.4 & 2.0 \\
\ce{NdCo_{1.50}Ni_{0.50}}   & 48 & 6.2 / 6.1 / - & 184  & 1.9 / 1.9 & - & - & - \\
\ce{NdCo_{1.25}Ni_{0.75}}   & 42 & 5.9 / 6.2 / - & 177  & 1.9 / 2.0 & - & - & - \\
\ce{NdCoNi}                 & 34 & 5.9 / 6.2 / 6.9 & 173  & 2.0 / 2.2 & 2.5 & 3.3 & 4.9 \\[1mm]
\end{tabular}
\end{ruledtabular}
\label{tab:mce}
\end{table*}